\documentclass[nonacm]{acmart}
\authorsaddresses{}

\AtBeginDocument{%
  }

\setcopyright{acmlicensed}
\copyrightyear{2018}
\acmYear{2018}
\acmDOI{XXXXXXX.XXXXXXX}
\acmConference[Conference acronym 'XX]{Make sure to enter the correct
  conference title from your rights confirmation email}{June 03--05,
  2018}{Woodstock, NY}
\acmISBN{978-1-4503-XXXX-X/2018/06}

\usepackage{blindtext}
\usepackage{graphicx}
\usepackage{subcaption}
\usepackage{enumitem}
\usepackage{comment}
\usepackage{array,booktabs,longtable}
\usepackage[normalem]{ulem} 
\usepackage[most]{tcolorbox}

\usepackage{tabularx}
\usepackage{microtype}
\usepackage{array}        

\newcommand{\SQ}[1]{\par\medskip\noindent\textbf{Q.}~#1\par\nobreak\smallskip}

\newcommand{\route}[1]{{\small\textcolor{gray}{[#1]}}}

\newenvironment{choices}%
  {\begin{list}{$\bigcirc$}{%
      \setlength{\leftmargin}{2em}\setlength{\itemsep}{0.15em}%
      \setlength{\topsep}{0.25em}\setlength{\parsep}{0pt}}}%
  {\end{list}}

\newenvironment{checks}%
  {\begin{list}{$\square$}{%
      \setlength{\leftmargin}{2em}\setlength{\itemsep}{0.15em}%
      \setlength{\topsep}{0.25em}\setlength{\parsep}{0pt}}}%
  {\end{list}}

\newcommand{\freetext}{\par\smallskip\noindent\rule{\linewidth}{0.4pt}\par\smallskip}

\newcommand{\blank}[1][6em]{\rule[-0.3ex]{#1}{0.4pt}}

\definecolor{cerulean}{rgb}{0.10, 0.58, 0.75}

\usepackage[capitalize]{cleveref}
\crefname{section}{Section}{Sections}
\Crefname{section}{Section}{Sections}
\Crefname{table}{Table}{Tables}
\crefname{table}{Table}{Tables}
\crefname{figure}{Figure}{Figures}

\usepackage{tikz}
\usetikzlibrary{positioning}
\definecolor{headerbg}{HTML}{F2F2F7}
\definecolor{purpleline}{HTML}{7B52C4}
\definecolor{purplefill}{HTML}{FBF9FE}
\definecolor{blueline}{HTML}{4E9AC6}
\definecolor{bluefill}{HTML}{F3FAFD}
\definecolor{tealline}{HTML}{86BDB4}
\definecolor{tealfill}{HTML}{F4FAF9}
\definecolor{orangeline}{HTML}{E0A44A}
\definecolor{orangefill}{HTML}{FDF5EA}
\definecolor{txtpurple}{HTML}{7030A0}
\definecolor{txtblue}{HTML}{3C9DD0}
\definecolor{txtgold}{HTML}{C0912C}

\begin{document}

\title[Use and Effects of LLMs in Peer Review: A Randomized Experiment and Survey at ICML 2026]{Use and Effects of LLMs in Peer Review:\\A Randomized Experiment and Survey at ICML 2026}

\author{Sunnie S. Y. Kim}
\authornote{Both authors contributed equally to this research. \\Corresponding authors: \{sunniesykim, wesleydeng, mdudik\}@microsoft.com\\}
\affiliation{%
  \institution{Microsoft Research}
  \city{New York}
  \country{USA}
}

\author{Wesley Hanwen Deng}
\authornotemark[1]
\affiliation{%
  \institution{Microsoft Research}
  \state{New York}
  \country{USA}
}

\author{Jennifer Wortman Vaughan}
\affiliation{%
  \institution{Microsoft Research}
  \state{New York}
  \country{USA}
}

\author{Buxin Su}
\affiliation{%
  \institution{University of Pennsylvania}
  \state{Philadelphia}
  \country{USA}
}

\author{Weijie Su}
\affiliation{%
  \institution{University of Pennsylvania}
  \state{Philadelphia}
  \country{USA}
}

\author{Alekh Agarwal}
\affiliation{%
  \institution{Google Research}
  \state{Kirkland}
  \country{USA}
}

\author{Sharon Li}
\affiliation{%
  \institution{University of Wisconsin-Madison}
  \state{Madison}
  \country{USA}
}

\author{Martin Jaggi}
\affiliation{%
  \institution{EPFL}
  \state{Lausanne}
  \country{Switzerland}
}

\author{Daniel G. Goldstein}
\affiliation{%
  \institution{Microsoft Research}
  \state{New York}
  \country{USA}
}

\author{Nihar B. Shah}
\affiliation{%
  \institution{Carnegie Mellon University}
  \state{Pittsburgh}
  \country{USA}
}

\author{Miroslav Dudík}
\affiliation{%
  \institution{Microsoft Research}
  \state{New York}
  \country{USA}
}

\renewcommand{\shortauthors}{Kim*, Deng* et al.}

\begin{abstract}
LLMs are rapidly reshaping peer review, making it important to understand how reviewers use them in practice and how different LLM-use policies affect review outcomes. We investigate these questions through a randomized experiment and an anonymous post-survey at ICML 2026, a major machine learning conference involving over 24,000 papers and 17,000 reviewers. Reviewers were assigned to either a conservative policy prohibiting all LLM use or a permissive policy allowing limited assistance, with randomization among a subset of main-track papers and reviewers. Policy assignment had near-zero effects on final paper decisions, paper scores, and reviewer confidence, although reviews under the permissive policy were 5.5-7\% longer. Post-survey responses (N=1,486) revealed diverse attitudes toward LLMs and substantial noncompliance: 22.5\% of conservative-policy reviewers reported using an LLM despite the prohibition, and 36.5\% of permissive-policy reviewers reported at least one explicitly disallowed use. We discuss implications for future peer-review policy and tool design. 
\end{abstract}

 


\maketitle

\section{Introduction}

Rapidly growing submission volumes are increasing the burden on peer-review systems and amplifying longstanding concerns about ``reviewer fatigue''~\cite{publons2018global}. At CHI and ICML, two leading conferences in human-computer interaction (HCI) and machine learning (ML), the increase in submissions between 2023 and 2026 was 3,182 to 6,730 (+112\%) and 6,538 to 24,661 (+277\%), respectively. In response to these pressures and the growing capabilities of large language models (LLMs), it has become common for researchers to incorporate the use of LLMs into their peer review practices~\cite{liang2024monitoring,liao2024llms, larosa2026ai, lemberger2026authors}.
Computer Science (CS) conferences, in turn, have begun to publish explicit peer-review LLM policies and explore the use of LLM assistance in the review workflow~\cite{thakkar2026large,icml2025policy, aistats2025policy, cvpr2026policy, eccv2026policy, iccv2025policy, aistats2026policy, chi2026policy, naturepolicy, arrpolicy, beygelzimer2023neurips}.\footnote{In Computer Science, conferences are a terminal publication venue: many conferences including ICML review full papers rather than just abstracts, and are often considered comparable to or more prestigious than journals.} (See \cref{sec:llmpolicies} for an overview of these policies.) \looseness=-1

Research to date has concentrated on what LLMs can and cannot do for peer review: exploring whether they can evaluate if the claims made in a paper are supported with evidence~\cite{liu2023reviewerGPT,xi2025flaws}, benchmarking the quality of model-generated reviews against human ones \cite{thakkar2025llmfeedback, baumann2026stop, chen2026happens}, probing models' failure modes such as sycophancy and hallucination \cite{sharma2026llms, liang2024monitoring}, and, more recently, designing and building LLM-powered tools to assist reviewers \cite{chen2025envisioning, zhu2026humanly}. 

However, we currently lack an empirical understanding of \textit{how peer-review LLM policies shape review outcomes}---a causal question best addressed through randomized controlled trials (RCTs) in live review settings rather than retrospective observational analyses (c.f., \cite{jecmen2020mitigating,saveski2023counterfactual}).
In addition, we know surprisingly little about \textit{how reviewers actually use LLMs in practice}---at which stage of their reviewing workflow, for which tasks, with how much verification, and where use shades into misuse---and \textit{whether these uses comply with conferences' LLM policies}. Understanding these questions is important for designing future policies and tools that are grounded in the realities of peer-review practice. \looseness=-1

In this work, we investigate these questions through a large-scale RCT and a post-survey conducted in collaboration with the organizers of the International Conference on Machine Learning (ICML) 2026, which  involved 76,159 authors, 17,886 reviewers, and 24,661 submitted papers. The RCT randomized a subset of submitted papers and reviewers across two policies for LLM use (see \cref{sec:policyinfo} for the full policy descriptions):
\begin{itemize}
    \item \textbf{Conservative Policy:} No LLM use was permitted during the review process.
    \item \textbf{Permissive Policy:} LLM use was permitted for understanding submitted papers and related work and for polishing reviewer-written text, but not for evaluating submitted papers or drafting reviews.
\end{itemize}
The study examined how policy assignment affects paper scores, final paper decisions, reviewer confidence, review length, and review quality ratings from area chairs (ACs) overseeing the review process. 
Across the two randomizations, we found that \textbf{assignments of the reviewers to the conservative versus the permissive policies had near-zero effects on final paper decisions, paper scores, and reviewer confidence}. However, we found small effect on the review length and perceived review quality, with reviews produced under the permissive policy slightly longer (5.5\%--7.7\%) and of slightly higher perceived quality than those produced under the conservative policy.

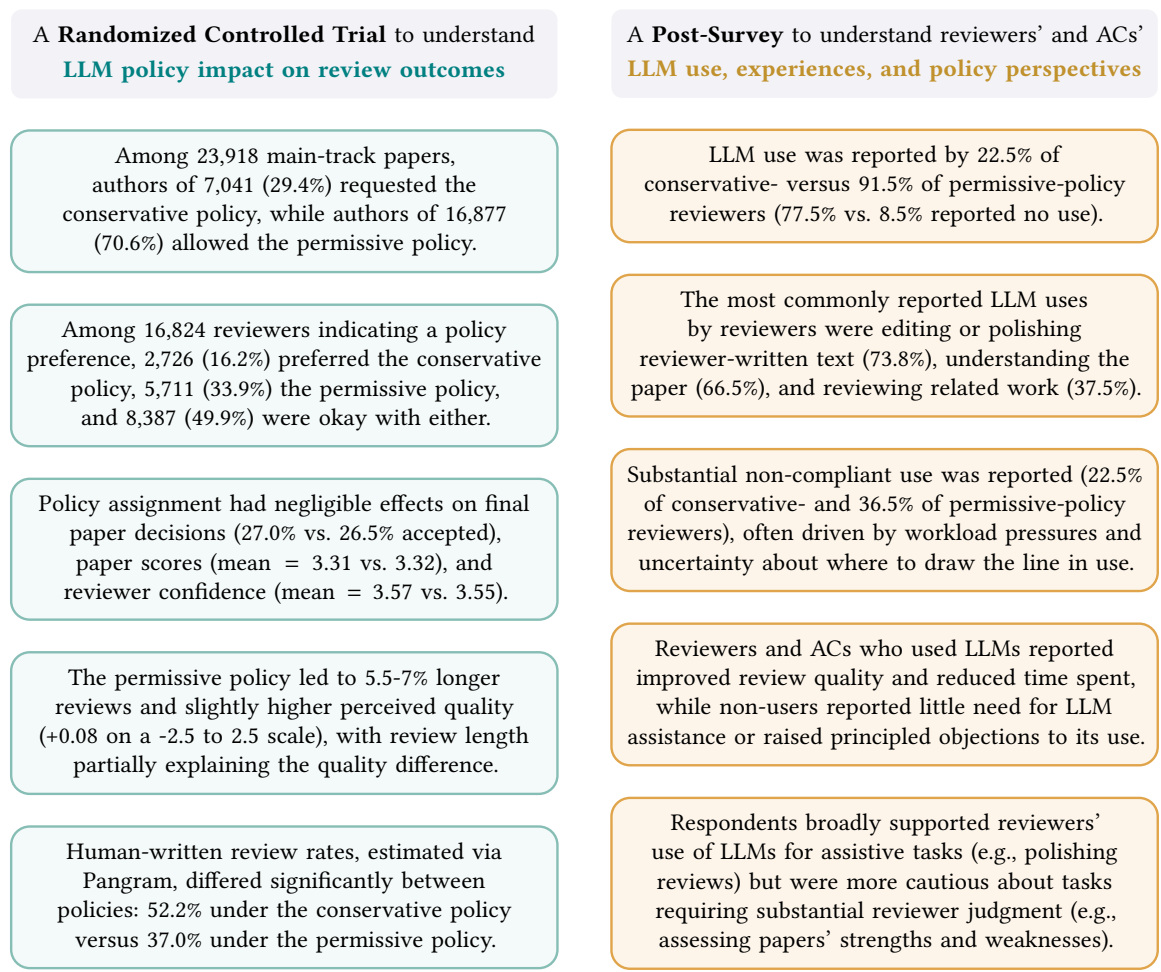
\begin{figure*}[h]
\centering
\begin{tikzpicture}[
font=\normalsize,
hdr/.style  = {rounded corners=4pt, fill=headerbg, align=center,
             inner xsep=6pt, inner ysep=6pt, anchor=north},
card/.style = {rounded corners=7pt, draw, line width=0.9pt, align=center,
             inner xsep=6pt, inner ysep=6pt, anchor=north},
purplecard/.style = {card, draw=purpleline, fill=purplefill},
bluecard/.style   = {card, draw=blueline,   fill=bluefill},
tealcard/.style   = {card, draw=tealline,   fill=tealfill},
orangecard/.style = {card, draw=orangeline, fill=orangefill},
node distance = 4mm and 0mm,
]

\newcommand{\LW}{0.45\linewidth}
\newcommand{\RW}{0.45\linewidth}
\node[hdr, text width=\LW] (LH) at (0,0)
  {A \textbf{Randomized Controlled Trial} to understand\\[2pt]
   \textcolor{teal}{\textbf{LLM policy impact on review outcomes}}};

\node[tealcard, text width=\LW, below=4mm of LH] (L1)
  {Among 23,918 main-track papers, authors of 7,041 (29.4\%) requested the conservative policy, while authors of 16,877 (70.6\%) allowed the permissive policy.};

\node[tealcard, text width=\LW, below=4mm of L1] (L2)
  {Among 16,824 reviewers indicating a policy preference, 2,726 (16.2\%) preferred the conservative policy, 5,711 (33.9\%) the permissive policy, and 8,387 (49.9\%) were okay with either.};

\node[tealcard, text width=\LW, below=4mm of L2] (L3)
  {Policy assignment had negligible effects on final paper decisions (27.0\% vs. 26.5\% accepted), paper scores ($\text{mean}=3.31$ vs. $3.32$), and reviewer confidence ($\text{mean}=3.57$ vs. $3.55$).
  };

\node[tealcard, text width=\LW, below=4mm of L3] (L4)
  {The permissive policy led to 5.5-7\% longer reviews and slightly higher perceived quality (+0.08 on a -2.5 to 2.5 scale), with review length partially explaining the quality difference.};

\node[tealcard, text width=\LW, below=4mm of L4] (L5)
  {Human-written review rates, estimated via Pangram, differed significantly between policies: 52.2\% under the conservative policy versus 37.0\% under the permissive policy.};

\node[hdr, text width=\RW, anchor=north west] (RH) at ([xshift=7mm]LH.north east)
  {A \textbf{Post-Survey} to understand reviewers' and ACs'\\[2pt]
  \textcolor{txtgold}{\textbf{LLM use, experiences, and policy perspectives}}};
  
\node[orangecard, text width=\RW, below=4mm of RH] (R1)
  {LLM use was reported by 22.5\% of conservative- versus 91.5\% of permissive-policy reviewers (77.5\% vs. 8.5\% reported no use).}; 
  
\node[orangecard, text width=\RW, below=4mm of R1] (R2)
  {The most commonly reported LLM uses by reviewers were editing or polishing reviewer-written text (73.8\%), understanding the paper (66.5\%), and reviewing related work (37.5\%).};
  
\node[orangecard, text width=\RW, below=4mm of R2] (R3)
  {Substantial non-compliant use was reported (22.5\% of conservative- and 36.5\% of permissive-policy reviewers), often driven by workload pressures and uncertainty about where to draw the line in use.};

\node[orangecard, text width=\RW, below=4mm of R3] (R4)
  {Reviewers and ACs who used LLMs reported improved review quality and reduced time spent, while non-users reported little need for LLM assistance or raised principled objections to its use.};

\node[orangecard, text width=\RW, below=4mm of R4] (R5)
  {Respondents broadly supported reviewers' use of LLMs for assistive tasks (e.g., polishing reviews) but were more cautious about tasks requiring substantial reviewer judgment (e.g., assessing papers' strengths and weaknesses).};

\end{tikzpicture}
\caption{\textbf{Summary of key findings from our randomized controlled trial (RCT) and post-survey (N=1{,}486).}}
\label{fig:summary}
\end{figure*}

After the review process concluded, we conducted an anonymous post-survey (N=1,486) of ICML 2026 reviewers, ACs, and senior ACs (SACs) to understand how reviewers actually used LLMs under their assigned policy, how LLM use shaped reviewers' experience, and respondents' perspectives on future LLM policies. 
We found substantial variation in experiences and attitudes toward LLMs: reviewers and ACs who used LLMs generally reported benefits such as improved review quality and reduced time spent, while many non-users reported little need for LLM assistance or raised principled objections to its use. We also observed substantial policy noncompliance: \textbf{22.5\% of reviewers under the conservative policy reported using an LLM despite the prohibition, and 36.5\% of reviewers under the permissive policy reported at least one explicitly disallowed use.} Finally, respondents broadly supported LLM use for assistive tasks but were more cautious about tasks requiring substantial reviewer judgment, while many questioned whether such boundaries could be effectively enforced. Key findings are summarized in \cref{fig:summary}.

Overall, this work provides a rare empirical look at LLM use in peer-review. In particular, it provides: \looseness=-1
\begin{itemize}
    \item Evidence from a large-scale RCT on how policies for LLM use in peer review affect review outcomes (\cref{sec:experiment});
    \item Insights from a large-scale post-survey on reviewers' and ACs' LLM use, experiences with LLMs, or reasons for non-use, as well as community perspectives on future LLM-use policies (\cref{sec:postsurvey});
    \item Implications for the design of peer-review policies and tools, informed by perspectives from reviewers, ACs, and SACs, many of whom were also authors of submissions at the same conference (\cref{sec:disc_policy});\looseness=-1
    \item Broader implications for how LLMs can support meaningful human engagement in peer review and beyond (\cref{sec:disc_engagement}).\looseness=-1
\end{itemize}

\section{Related Work}

This paper draws on and contributes to several bodies of research. In \cref{rw:human}, we review related research on LLM-assisted knowledge work, focusing on questions of human judgment, expertise, ownership, and accountability that are particularly relevant to peer review. In \cref{rw:empirical}, we then review empirical studies of LLM use and its effects on peer review. We also provide an overview of LLM policies for peer review at CS conferences in \cref{sec:llmpolicies}. \looseness=-1

\subsection{Human Judgment, Expertise, Ownership, and Accountability in LLM-Assisted Knowledge Work} \label{rw:human}

Peer review is a type of knowledge work in which researchers evaluate the correctness, novelty, and significance of research produced by their peers. It requires substantial expertise and is a high-stakes activity that shapes which research is recognized and disseminated, while influencing researchers' careers, funding, and future opportunities. As such, the use of LLMs in peer review raises questions about how LLM assistance affects reviewers' judgment, expertise, ownership, and accountability, as well as how knowledge workers comply with institutional policies governing LLM use.\looseness=-1

Concerns about preserving human judgment and expertise are particularly salient in peer review, where independent judgment and reviewer expertise are central. These concerns are supported by prior work in other domains suggesting that AI assistance can bias human judgments \cite{Jakesch2023Cowriting}, encourage (over)reliance on AI outputs \cite{kim2024uncertainty,kim2025fostering,ibrahim2026aies}, and reduce opportunities to develop expertise and critical-thinking skills \cite{lee2025impact,liu2026persistence,gerlich2025ai,zhi2026investigating,shen2026aiimpactsskillformation}. At a collective level, AI assistance may also lead to greater homogeneity in ideas and outputs \cite{anderson2024homogenization, padmakumar2024does,doshi2024generative,ashkinaze2025how}. At the same time, other work suggests that AI assistance can, in some settings, help people make better judgments \cite{shin2023pnas}, or support learning and expertise development \cite{wang2024a,Hsu2025Counseling,Nie2025GPTSurprise}. However, these potential benefits and risks are not well understood in the context of peer review.

Research also suggests that AI assistance can shape people's sense of ownership and accountability in nuanced ways. AI assistance generally reduces people's sense of ownership, although the effect can depend on factors such as when AI enters the process \cite{gero2026planning}. People may also experience an ``AI ghostwriter" effect, in which they do not consider themselves the authors of AI-generated text while also refraining from attributing authorship to AI \cite{draxler2024ghostwriter}. AI involvement may also shift accountability from people to the AI system, potentially resulting in lower-quality work or undesirable practices \cite{Sadeghian2024SoulofWork,wiles2026putting}. These findings motivated us to explore reviewers' perceptions of how LLM use affects ownership and accountability in our post-survey.

Finally, organizations are increasingly developing policies to govern AI use, but actual AI use often diverges from policy \cite{hwang2026forgiveness, wagman2025generative, gillespie2025trust, jeong2026everyday}. For example, roughly half of employees in a 48,340-respondent study reported using AI in ways that violate organizational policy, and over half reported concealing AI use and presenting AI output as their own \cite{gillespie2025trust}. Another study finds 
workers preferred a commercial AI tool over sanctioned internal tools and had inconsistent understandings of policy requirements \cite{wagman2025generative}. Policy compliance is also a challenge in peer review. At ICML 2026, submissions by reviewers 
found to violate the conference's LLM-use policy were desk-rejected~\cite{icml2026deskreject}, and our Pangram analysis and post-survey likewise suggested substantial noncompliance. \looseness=-1

This literature highlights recurring tensions surrounding LLM assistance in knowledge work, motivating the measures in our work. Our RCT examines how LLM-use policies affect review quality, reviewer confidence, review length, review scores, and paper decisions (\cref{fig:experiment1}). Our post-survey examines reviewers' perceived time, effort, review quality, confidence, engagement, and ownership (\cref{fig:reviewer-perceptions}), as well as verification practices, accountability for review content (\cref{fig:careful_standby}), and policy compliance (\cref{fig:llm-use}). In the next section, we provide more context on relevant empirical studies. \looseness=-1

\subsection{Empirical Studies of LLM Use in Peer Review} \label{rw:empirical}

There is a long tradition of CS conferences running experiments to understand and improve the review process \cite{cortes2021inconsistency,beygelzimer2023neurips}. Within the CHI community, researchers like Nobarany et al. have conducted surveys to understand reviewers' motivations for review articles \cite{nobarany2016motivates}. In recent years, a growing body of work explores LLMs' effects on peer review and their prevalence in conference review pipelines \cite{mann2025ai, liang2024monitoring, sharma2026llms, larosa2026ai}. For example, using a distributional detection method, \citet{liang2024monitoring} estimated that 6.5--16.9\% of review text across several major CS conferences could have been substantially modified by LLMs. While these works establish that LLM use is widespread, they \textit{infer} reviewer behavior from review artifacts \textit{retroactively} after reviews are submitted, rather than conducting RCTs during a live review process to isolate the causal impact of LLM usage. \looseness=-1

A second line of work evaluates the ability of LLMs to assist with peer review, particularly by identifying flaws in submissions. Early work found that LLMs could identify errors in short papers \cite{liu2023reviewerGPT}, while an experiment at a major ML conference found that only one of 79 human reviews identified an inserted flaw, whereas GPT-4 could identify certain flaws~\cite[Section 10.1.2]{shah2022acm}. More recent work has evaluated this capability at larger scale~\cite{xi2025flaws}, and LLMs have been deployed for flaw detection in the review process at some venues (e.g., TMLR)~\cite{rao2026evaluating,TMLRaireviews2026}. 
 
A third line of work designs and evaluates LLM support for reviewers. In a randomized study of over 20,000 ICLR 2025 reviews, \citet{thakkar2026large} found that 27\% of reviewers who received automated feedback revised their reviews, yielding longer and, under blinded evaluation, more informative reviews. However, follow-up surveys and interviews with these reviewers surfaced tensions between LLM-enabled efficiency gains and concerns about fairness and accountability \cite{chen2026happens}. \looseness=-1

A fourth line of work characterizes broader stakeholder attitudes toward AI assistance in research and peer review \cite{liao2024llms, lemberger2026authors, chen2025envisioning, rao2025ml}. In a survey of over 2,000 ML researchers, participants responded that open peer-review policies (such as those at ICML 2026 which release reviews of accepted papers) can help deter ``AI slop'' submissions and AI-generated reviews, while opinions were split on whether conferences should generate official AI reviews~\cite{rao2025ml}. In a user study on ChatGPT-assisted reviewing, \citet{chen2025envisioning} explored how reviewers integrate LLM outputs into their evaluation workflows and surfaced trade-offs between speed and critical thinking. 

Finally, a small but growing body of work examines the enforceability of LLM-use policies at major conferences. Evaluating five detectors on 50,156 reviews, \citet{saha2026policies} showed that human-written but LLM-polished reviews were frequently characterized as LLM-written by LLM-text-detectors, so that even low false-positive rates would produce hundreds of wrongful accusations at conference scale. Others argue that prohibition-based policies mainly create incentives to conceal \cite{larosa2026ai}. ICML 2026 itself employed prompt injections and identified several hundred reviews that had been fed to LLMs in violation of one of its policies~\cite{icml2026deskreject,Rao2025watermarking}. \looseness=-1

What remains missing is \textbf{\textit{direct empirical evidence} of how LLM policies shape review outcomes, how reviewers actually use LLMs under different policies, and what policies the community prefers.} We address this gap through a large-scale RCT in a live peer-review process and an anonymous survey of reviewers, ACs, and SAC respondents. These studies allow us to measure policy effects on review outcomes, examine LLM use and compliance, and explore community preferences for future LLM use and policies in peer review. \looseness=-1

\section{Research Setting: ICML 2026} \label{method: study context}

We next describe our research setting, ICML 2026, and provide context for our RCT and post-survey studies. We begin with an overview of the conference and its review process (\cref{sec:conferenceinfo}), then describe its dual-policy framework for reviewer LLM use (\cref{sec:policyinfo}), and finally go through the details of policy assignment and enforcement (\cref{sec:policyassignment}).

\subsection{Conference and Review Process}
\label{sec:conferenceinfo}

The International Conference on Machine Learning (ICML) is one of the largest ML conferences. ICML 2026, the 43rd edition of the conference, was held in Seoul, South Korea, from July 6 to 11, 2026, and featured two submission tracks. The \textit{main track} invited original research contributions of broad interest to the ML community, emphasizing novel methods, empirical findings, and theoretical advances~\cite{icml2026cfp}. The \textit{position track} invited papers that articulated viewpoints or perspectives on what should be done in the field of ML, generally taking a meta-level perspective rather than presenting new technical results~\cite{icml2026cfpposition}. Overall, ICML 2026 received 24,661 submissions from 76,159 authors, including 23,918 main-track submissions and 743 position-track submissions. Of these submissions, 6,552 (26.6\%) were accepted and 18,109 (73.4\%) were rejected or withdrawn. \looseness=-1

ICML 2026's program committee comprised 17,886 reviewers, 1,728 area chairs (ACs), 251 senior ACs (SACs), 56 ethics reviewers, 2 position-track chairs, and 4 program chairs (PCs). Note that the main track and the position track maintained separate reviewer and AC pools. In the paper review process, \textit{reviewers} read and evaluated assigned submissions and recommended decisions to ACs. This included reviewers recruited by PCs ahead of the conference, \textit{reciprocal reviewers} (qualified authors of submissions who were required to serve as reviewers as part of the conference's reciprocal reviewing system), and additional reviewers directly invited by ACs during the reviewing period to review specific papers. \looseness=-1

\textit{ACs} coordinated the review process for individual submissions by recruiting and managing reviewers, monitoring review quality, facilitating discussions, writing meta-reviews, and making recommendations. In main track, \textit{SACs} supervised ACs, discussed borderline papers, promoted consistency across decisions, and supported the PCs in making the final decisions. In the position track, position-track chairs supervised ACs and made the final decisions. Separately, \textit{ethics reviewers} provided 
assessments of submissions flagged for potential ethical concerns. \textit{PCs} oversaw the entire peer-review process and the conference program.\looseness=-1

\subsection{Dual-Policy Framework for LLM Use in Reviewing}
\label{sec:policyinfo}

While the prior year's conference (ICML 2025) prohibited reviewer LLM use altogether, ICML 2026 introduced a dual-policy framework governing LLM use in the main-track review process~\cite{icml2026policy}, informed by two community surveys of ICML 2025 reviewers conducted in November 2025 ($N$=150 and $N$=74)~\cite{icml2026policyintro}. Position-track submissions were reviewed by a separate reviewer pool whose members were all prohibited from using LLMs. Accordingly, unless otherwise noted, all subsequent descriptions refer to main-track papers and reviewers.

The dual-policy framework comprised a \textbf{conservative policy (C)} and a \textbf{permissive policy (P)}. The policy descriptions provided to authors and reviewers are reproduced in \cref{fig:policy}.

\begin{figure}[h]
\begin{tcolorbox}[title=ICML 2026 Policy for LLM Use in Reviewing]

    \textbf{Conservative Policy}: Use of LLMs in any stage of reviewing is \uline{strictly prohibited} (except for inadvertent use in tools that are not traditionally LLM-based, like web search/retrieval and spelling/grammar checkers).
    \\[1.2em]
    \textbf{Permissive Policy}:
    \\[0.5em]
    \textit{\textbf{Allowed:}} \uline{Use of LLMs to help understand the paper and related works, and polish reviews.} Submissions can be fed to privacy-compliant* LLMs. 
    \\[0.5em]
    \textit{*By ``privacy-compliant", we refer to LLM tools that do not use logged data for training and that place limits on data retention. This includes enterprise/institutional subscriptions to LLM APIs, consumer subscriptions with an explicit opt-out from training, and self-hosted LLMs. (We understand that this is an oversimplification.)}
    \\[0.5em]
    \textit{\textbf{Not allowed:}} \uline{Delegation of judgement and critique of the paper to LLMs.} This includes asking LLMs to assess paper’s quality/significance, identify paper’s strengths/weaknesses, suggest key points for the review, suggest an outline for the review, write the full review, or suggest questions for authors.
    \\[1.2em]
    Under both policies, reviewers are expected to read the entire body of the paper, but not necessarily appendices and supplementary material. \uline{Reviewers are fully responsible} for the entire content of their reviews, regardless of the AI tools that they use. In particular, any hallucinated content in reviews is subject to disciplinary action according to our academic integrity policy. Low-quality reviews may result in penalties according to our reciprocal reviewing policy.
 
\end{tcolorbox}
\caption{\textbf{ICML 2026 policy descriptions provided to authors and reviewers (reproduced from \cite{icml2026policy}).}}
\label{fig:policy}
\end{figure}

These two policies are broadly consistent with policies adopted at other major CS conferences reviewed in \cref{sec:llmpolicies}. Most conferences either prohibit reviewer LLM use altogether (similar to ICML 2026's conservative policy) or permit only limited uses, such as assistance with polishing writing or understanding submitted papers (similar to ICML 2026's permissive policy). What distinguishes ICML 2026 is its dual-policy framework, which accommodates author and reviewer preferences, and its controlled experiment examining the effects of the two policies.

\subsection{Policy Assignment and Enforcement}
\label{sec:policyassignment}

Prior to reviewer assignment, both authors and reviewers declared their policy preferences. Authors could require their paper to be reviewed under the conservative policy or could indicate that their paper could be reviewed under either policy, with the understanding that requiring the conservative policy for a paper would also commit all authors of the paper who are also reviewers to be willing to review under the conservative policy. Reviewers indicated whether they preferred the conservative policy, preferred the permissive policy, or were okay with either. \looseness=-1 

\begin{figure*}[t]
  \centering
  \includegraphics[width=\linewidth]{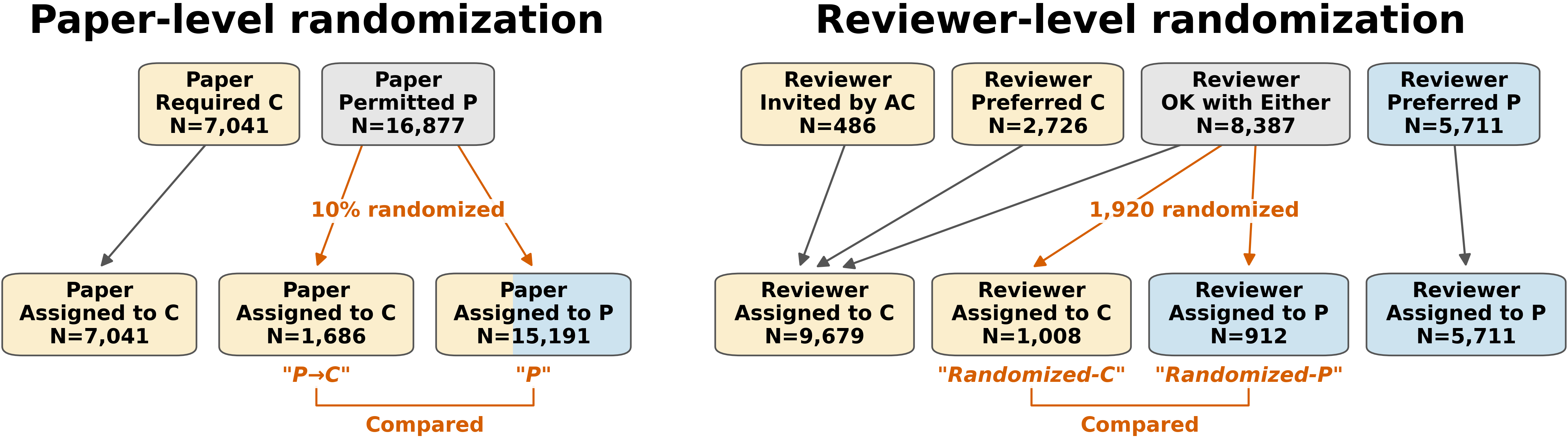}
    \caption{\textbf{Overview of the RCT design.} Yellow and blue indicate the permissive (P) policy and the conservative (C) policy, respectively. For papers, colors indicate the policy of the reviews they receive; for reviewers, the policy of the reviews they write. Papers assigned to the permissive policy are shown in both colors because they can receive reviews from reviewers assigned to either policy. Comparisons are restricted to the randomized subsets shown at the bottom of each panel. The RCT included only main-track papers and reviewers.}
\label{fig:experimentdesign}
\vspace{0.3cm}
  \centering
  \includegraphics[width=\linewidth]{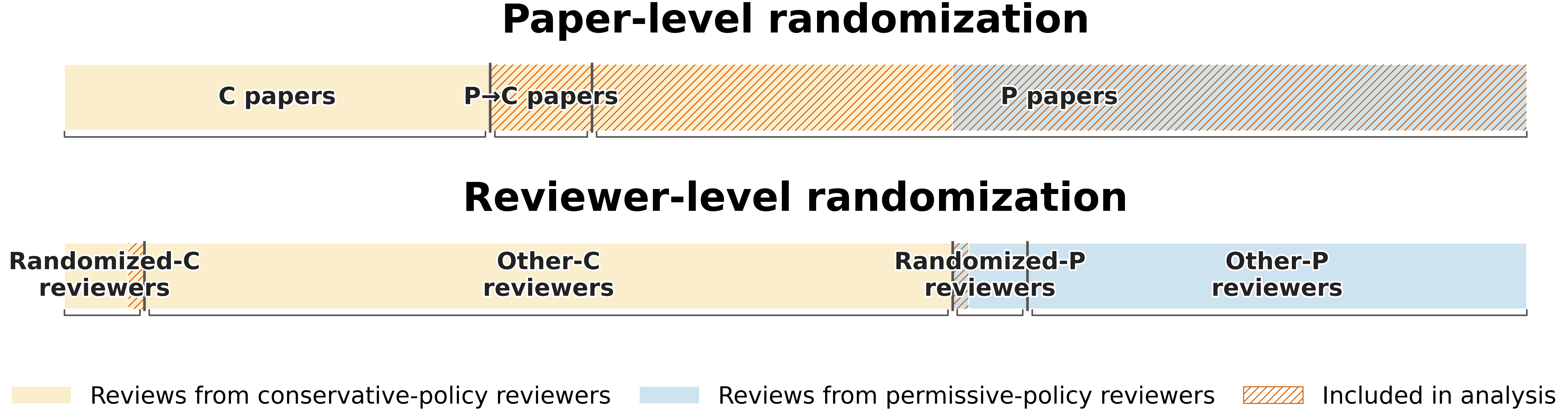}
  \caption{\textbf{Distribution of 90,822 main-track reviews across paper-level and reviewer-level policy assignments in the RCT.} Hatched regions indicate reviews included in the randomized comparisons: reviews for P$\rightarrow$C versus P papers in the paper-level analysis, and paired subsets of reviews from Randomized-C versus Randomized-P reviewers in the reviewer-level analysis.}
\label{fig:experiment2_design}
\end{figure*}

As shown in \cref{fig:experimentdesign}, among the 23,918 main-track papers, authors of 7,041 (29.4\%) papers required the conservative policy for their papers, while authors of 16,877 (70.6\%) papers indicated their papers could be reviewed under either policy. 
Among the 17,310 main-track reviewers, 486 were invited by ACs during the reviewing period and were required to follow the conservative policy. The remaining 16,824 reviewers were asked for their policy preference before the reviewing period. Among them, 2,726 (16.2\%) preferred the conservative policy, 5,711 (33.9\%) preferred the permissive policy, and 8,387 (49.9\%) indicated that they were okay with either policy. \looseness=-1

Submissions were matched with reviewers whose policy preferences were compatible with author requirements, and each reviewer was assigned a single policy that governed all of their reviews. Policy assignments were communicated through multiple channels. In particular, assigned policies were displayed in the OpenReview reviewer console, and reminders were shown on every submission assigned to each reviewer. See \cref{fig:reviewerconsole,fig:reminder} for screenshots.

The conference implemented a mechanism to detect potential violations of the conservative policy \cite{Rao2025watermarking,icml2026deskreject}. Specifically, each submission being reviewed under the conservative policy was distributed as a uniquely modified PDF containing hidden machine-readable instructions. If a reviewer uploaded the submission to an LLM during the review process, the instructions directed the LLM to include two specific phrases, randomly selected from a dictionary of approximately 170,000 phrases, in its output. Reviews were automatically screened for these phrases, and every flagged case was manually verified to rule out benign explanations, such as a reviewer noticing and mentioning the embedded instructions.

The appearance of both embedded phrases in a review therefore provided strong evidence that a submission had been shared with an LLM during the review process, in violation of the conservative policy. According to the ICML blog post~\cite{icml2026deskreject}, the watermarking scheme was designed such that the probability of even a single false positive across all roughly 53,000 reviews under the conservative policy, that is, the family-wise error rate, was approximately 1 in 10,000. \looseness=-1

In total, 398 reciprocal reviewers were found to have violated the conservative policy, resulting in 497 desk rejections of their submissions.
Including reviews that were not by reciprocal reviewers, around 1.5\% of the roughly 53,000 reviews under the conservative policy were flagged; for comparison, the same test flagged around 2.9\% of the roughly 34,000 reviews under the permissive policy, but these did not result into any enforcement action.
However, our Pangram analysis of a subset of reviews (\cref{sec:pangram}) and post-survey responses (\cref{sec:postsurveyresults}) suggest substantially higher rates of policy noncompliance. This difference is unsurprising because the watermarking mechanism was designed to detect only a specific subset of violations. Nevertheless, this gap highlights the practical challenges of enforcing LLM-use policies, a topic we return to later in the paper (\cref{sec:discussion}).

\section{RCT on the Effects of LLM Use on Review Outcomes}
\label{sec:experiment}

\subsection{RCT Methods} \label{Experiment Methods}

To examine the effects of the two LLM-use policies, we conducted a large-scale RCT on the ICML 2026 review process in collaboration with the conference organizers. The experiment spanned the period from full-paper submission (January 28, 2026) to author notification (April 30, 2026), and was approved by the University of Pennsylvania's Institutional Review Board (IRB).
The experiment involved randomization at both the paper and reviewer levels. At a high level, the paper-level randomization was designed to help answer the question \textit{``Do review outcomes differ when a paper is reviewed under the conservative policy versus the permissive policy?"} In contrast, the reviewer-level randomization was designed to help answer the question \textit{``Do reviewers assigned to the conservative policy and the permissive policy produce different reviews for the same paper?"} Randomized assignments were implemented within the policy-assignment mechanism described in \cref{method: study context} and did not override any stated author requirements or reviewer preferences. \cref{fig:experimentdesign,fig:experiment2_design} provide an overview of the experimental design. \looseness=-1

\subsubsection{Paper-Level Randomization} 
To estimate the effects of the two policies while respecting author choice, paper-level randomization was conducted among main-track papers whose authors permitted review under the permissive policy ($N=16{,}877$). Specifically, 10\% of these papers were randomly switched to the conservative policy (P$\rightarrow$C; $N=1{,}686$), while the remainder stayed under the permissive policy (P;  $N=15{,}191$). Note that papers assigned to the conservative policy were reviewed exclusively by reviewers assigned to the conservative policy. In contrast, papers assigned to the permissive policy could be reviewed by reviewers assigned to either policy. \looseness=-1

The primary outcomes were final paper decision (rejected/withdrawn, regular acceptance, or spotlight acceptance) and four paper-level review outcomes: mean paper score (range = 1--6), mean reviewer confidence (range = 1--5), mean review length (characters), and mean standardized AC quality rating (range = -2.5--2.5). AC quality ratings were originally collected on a 0--6 scale. To account for differences in rating tendencies across ACs, ratings were standardized within AC as $z$-scores and clipped to $[-2.5, 2.5]$, before being averaged at the paper level. See \cref{sec:reviewform,sec:acreviewquality} for the relevant review fields and questions. Strictly speaking, paper score, reviewer confidence, and the original AC quality ratings are ordinal data, but we treated their adjacent scale points as approximately equally spaced for these analyses. \looseness=-1

For final paper decisions, we compared distributions using a Pearson $\chi^2$ test of homogeneity. For the other four numeric outcomes, we compared means using Welch's independent-samples $t$-tests and compared full distributions using two-sample Kolmogorov-Smirnov tests with 10,000 random permutations. As a robustness check for the Welch's tests, we also ran studentized randomization tests, which produced identical $p$-values to two decimal places. \looseness=-1

\subsubsection{\mbox{Reviewer-Level Randomization}}
We randomized reviewers after the initial paper-reviewer assignment.
To be part of the randomized cohort, reviewers had to satisfy two conditions: (1) they indicated they were willing to review under either policy ($N=8{,}387$); (2) all papers assigned to them allowed the permissive policy. Reviewers satisfying these two conditions were randomly assigned either to the conservative policy (Randomized-C; $N=1{,}008$) or to the permissive policy (Randomized-P; $N=912$).\looseness=-1

This design ensures that policy assignment is independent of reviewer characteristics, allowing us to evaluate the impact of the assigned policy on the randomized cohort. However, reviewers eligible for randomization are not a representative subset of all reviewers. The two eligibility conditions may be correlated with factors such as reviewers' fields of expertise and seniority; as such, results from this comparison should be interpreted with this limitation in mind.\looseness=-1

To control for differences in the papers reviewed by the randomized cohort, we restricted the analysis to within-paper comparisons. Specifically, we identified 940 papers that received at least one review from a Randomized-C reviewer and at least one review from a Randomized-P reviewer. This design allows us to compare reviews written under the two randomized policy conditions for the same paper, thereby holding paper characteristics constant. Since Randomized-P reviewers were eligible to review only permissive-policy papers, all papers included in this analysis were permissive-policy papers.\looseness=-1

The primary outcomes were paper score (range = 1--6), reviewer confidence (range = 1--5), review length (characters), and standardized AC quality ratings (range = -2.5--2.5). As above, we treated paper score, reviewer confidence, and the original AC quality ratings as having approximately equally spaced scale points. The unit of analysis was the paper. For each outcome, we computed the mean among Randomized-C reviews and the mean among Randomized-P reviews within each paper. We then tested whether the mean within-paper difference between conditions differed from zero using two-sided paired $t$-tests.
\looseness=-1

\subsubsection{Pangram Analysis of a Subset of Reviews} 
As a supplementary analysis, we ran a subset of reviews through Pangram \cite{pangram2026}, a classifier of AI-generated text, to assess AI involvement across 
policy groups. Of the 90,822 main-track reviews, we first sampled 800 reviews for the paper-level randomization, with 100 reviews from each of the eight feasible combinations of paper and reviewer policy choice/assignment.\footnote{There are three paper policy assignment groups (C$\rightarrow$C, P$\rightarrow$C, and P) and four reviewer policy choice-to-assignment groups (C$\rightarrow$C, Either$\rightarrow$C, Either$\rightarrow$P, and P$\rightarrow$P). This yields eight possible combinations because Either$\rightarrow$P and P$\rightarrow$P reviewers can review only P papers, whereas C$\rightarrow$C and Either$\rightarrow$C reviewers can review papers under any policy assignment.}  We then sampled additional reviews for the reviewer-level randomization to obtain 500 reviews from each of the Randomized-P and Randomized-C groups. We submitted the full text of each review to the Pangram API. Among the fields returned by the API, we focused on \texttt{prediction\_short}, which classifies each review as human-written, AI-generated, or mixed. \looseness=-1

\begin{figure*}[!htbp]
    \centering
    \includegraphics[width=\linewidth]{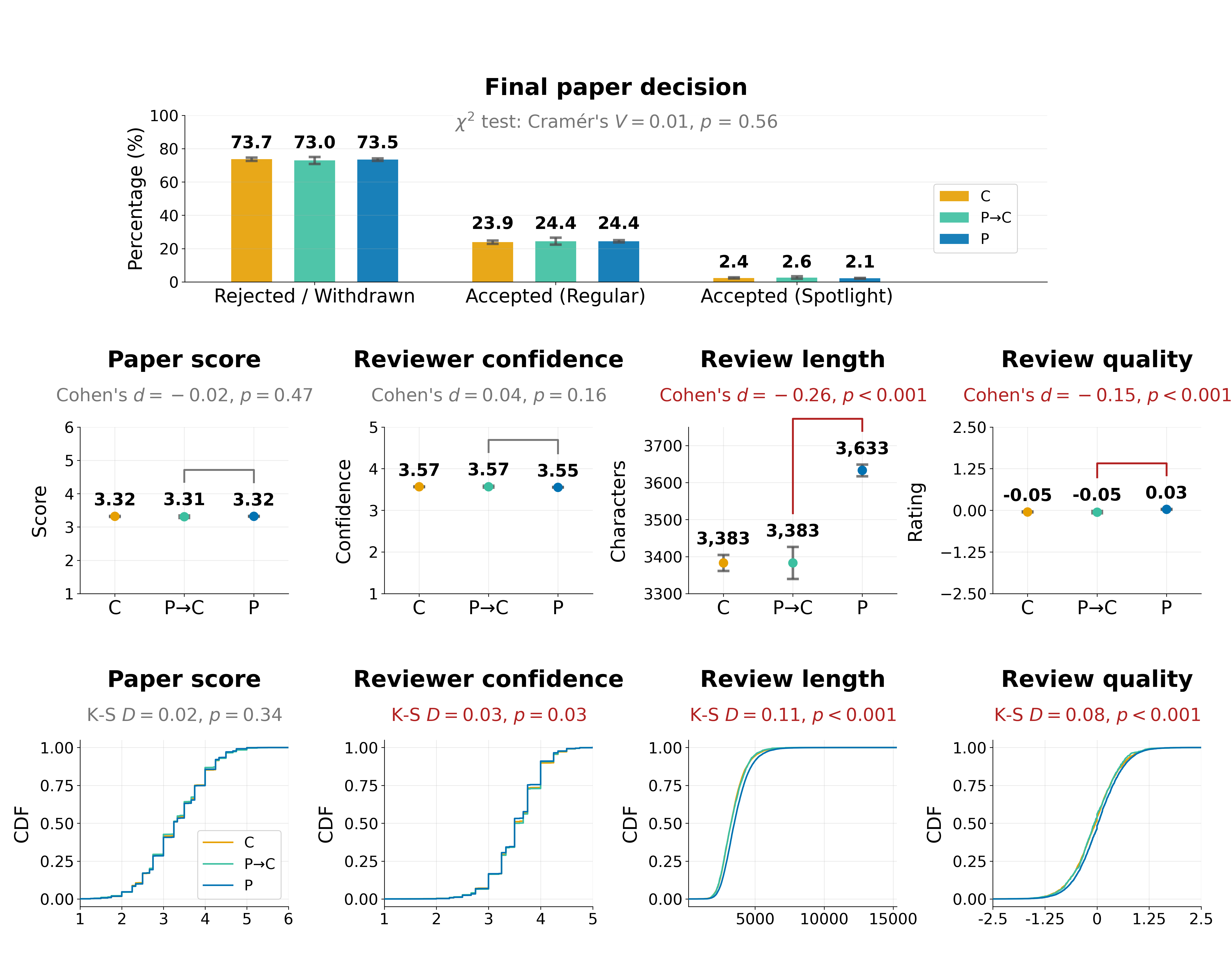}
    \caption{\textbf{RCT paper-level randomization results} comparing P$\rightarrow$C and P group papers. The top row shows final paper decisions by policy group, with effect size and $p$-value from a Pearson $\chi^2$ test of homogeneity. For numeric outcomes, the middle row shows group means with 95\% confidence intervals, effect sizes, and $p$-values from Welch's independent-samples $t$-tests; the bottom row shows empirical cumulative distribution functions (CDFs), effect sizes, and $p$-values from Kolmogorov--Smirnov (K--S) tests.\looseness=-1}
    \label{fig:experiment1}
  \centering
  \includegraphics[width=\linewidth]{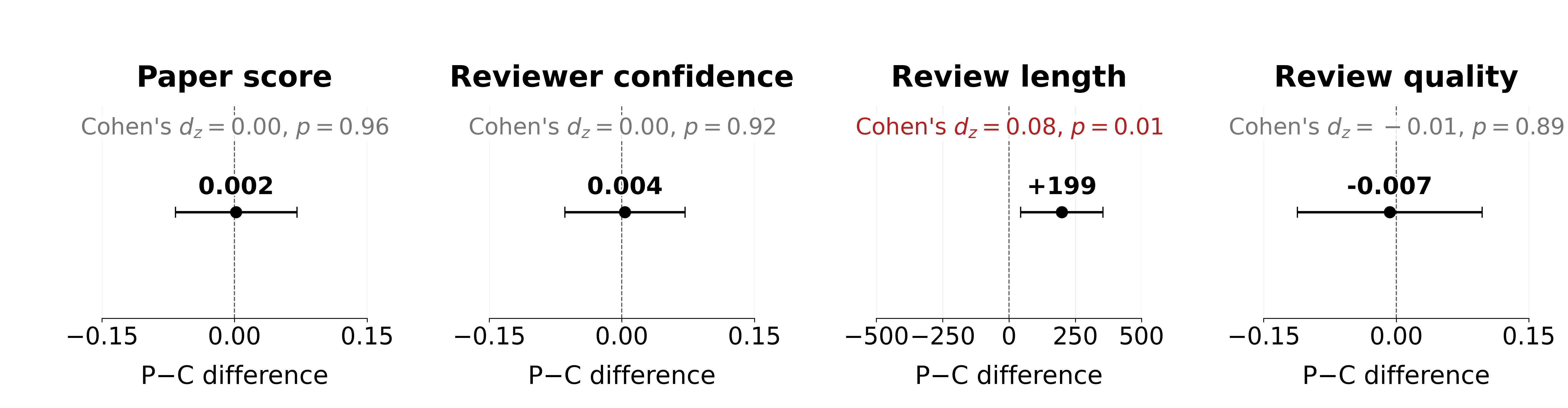}
  \caption{\textbf{RCT reviewer-level randomization results} comparing Randomized-P and Randomized-C reviews among papers that received reviews from both groups. Points show estimated mean within-paper differences with 95\% confidence intervals, effect sizes, and $p$-values from two-sided paired $t$-tests.}
  \label{fig:experiment2_effects}
\end{figure*}

\subsection{RCT Results}
\label{sec:rctresults}

\subsubsection{Paper-Level Randomization Results}

As described above, we evaluated the effect of policy assignment by randomly selecting 10\% of papers that permitted the permissive policy, reassigning them to the conservative policy (P$\rightarrow$C; $N=1{,}686$), and comparing them with papers retained under the permissive policy (P; $N=15{,}191$). Figure~\ref{fig:experiment1} summarizes the results. C is included as a reference point, but all statistical comparisons were between P$\rightarrow$C and P. \looseness=-1

\textbf{Policy assignment had near-zero effects on final paper decisions, paper scores, and reviewer confidence.}
Final paper decisions were nearly identical between P$\rightarrow$C and P ($\chi^2(2)=1.18$, Cramér's $V=0.01$, $p=0.56$). In P$\rightarrow$C and P, respectively, 73.0\% and 73.5\% of papers were rejected or withdrawn, 24.4\% and 24.4\% were accepted as regular papers, and 2.6\% and 2.1\% were accepted as spotlights.

Mean paper scores (1--6) were also nearly identical for papers in P$\rightarrow$C ($M=3.31$, $SD=0.74$) and P ($M=3.32$, $SD=0.72$). Neither the difference in means ($t(2048.05)=-0.73$, Cohen's $d=-0.02$, $p=0.47$) nor the difference in score distributions (K--S $D=0.02$, $p=0.34$) was statistically significant. 

Similarly, mean reviewer confidence was nearly identical for papers in P$\rightarrow$C ($M=3.57$, $SD=0.47$) and P ($M=3.55$, $SD=0.46$), and the difference in means was not statistically significant ($t(2041.49)=1.39$, Cohen's $d=0.04$, $p=0.16$). Although the K--S test detected a difference between the distributions (K--S $D=0.03$, $p=0.03$), the effect size was very small and the empirical cumulative distribution functions (CDFs) largely overlapped.

\textbf{Papers reviewed under the permissive policy received slightly longer and higher-rated reviews.}
Papers in P received slightly longer reviews than papers in P$\rightarrow$C ($M=3{,}633$, $SD=982$ versus $M=3{,}383$, $SD=897$), corresponding to a mean difference of 250 characters (95\% CI $[204,296]$), or approximately 7\% longer reviews. The effect size was small (Cohen's $d=0.26$; $t(2142.95)=10.71$, $p<0.001$), as was the difference between the review-length distributions (K--S $D=0.11$, $p<0.001$). \looseness=-1

Papers in P likewise received reviews with slightly higher standardized AC quality ratings than papers in P$\rightarrow$C ($M=0.03$, $SD=0.54$ versus $M=-0.05$, $SD=0.54$), corresponding to a mean difference of 0.08 points on the standardized -2.5 to 2.5 rating scale (95\% CI $[0.05,0.12]$). However, the effect size was again small (Cohen's $d=0.15$; $t(1234.31)=4.62$, $p<0.001$), as was the difference between the quality-rating distributions (K--S $D=0.08$, $p<0.001$).

\textbf{Additional analyses.} Recall that while conservative-policy papers were reviewed exclusively by conservative-policy reviewers, permissive-policy papers could be reviewed by a mix of permissive- and conservative-policy reviewers. As a robustness check, we examined outcomes by the percentage of reviewers assigned to the permissive policy (0\%, 25\%, 50\%, 75\%, or 100\%) and found similar results across groups (\cref{fig:experiment1_robustness}).

Second, inspired by \citet{goldberg2025} (NeurIPS 2022 experiment), who found that evaluators gave higher quality ratings to reviews artificially lengthened without adding useful information, we examined whether review length was associated with review quality ratings using a regression analysis. A 1,000-character increase in mean review length was associated with a 0.20-unit increase in standardized AC quality ratings (95\% CI $[0.19,0.21]$, $p<0.001$). Adjusting for review length reduced the quality-rating difference between P and P$\rightarrow$C from $0.08$ to $0.04$, suggesting that longer reviews may partially explain the higher quality ratings under P. However, this is an exploratory analysis and does not establish a causal effect of review length on perceived review quality, unlike the experimental design in \cite{goldberg2025}.

\subsubsection{Reviewer-Level Randomization Results}

Here, as described above, we evaluated the effect of policy assignment by comparing reviews written by reviewers randomized to the conservative policy (Randomized-C) and the permissive policy (Randomized-P) within the 940 papers that received reviews from both groups. Figure~\ref{fig:experiment2_effects} summarizes the results.

\textbf{Reviewers assigned to the permissive policy wrote slightly longer reviews, but policy assignment had little effect on paper scores, reviewer confidence, or review quality.} 
Across 940 paired papers, Randomized-P reviews were, on average, 199 characters longer than Randomized-C reviews (P$-$C $=199$, 95\% CI $[44,354]$, Cohen's $d_z=0.08$, $p=0.01$). Although statistically significant, the effect size was very small.

In contrast, paper scores were nearly identical between groups (P$-$C $=0.002$, 95\% CI $[-0.067,0.071]$, Cohen's $d_z=0.00$, $p=0.96$), as was reviewer confidence (P$-$C $=0.004$, 95\% CI $[-0.064,0.072]$, Cohen's $d_z=0.00$, $p=0.92$). The estimated differences were near zero, and effect sizes were negligible for both outcomes.

Among the 560 paired papers with AC quality ratings available under both groups, standardized AC quality ratings likewise did not differ between groups (P$-$C $=-0.007$, 95\% CI $[-0.112,0.097]$, Cohen's $d_z=-0.01$, $p=0.89$). The estimated difference was near zero, and the effect size was negligible.

\textbf{Results were similar across the two randomizations.}
In both the paper-level and the reviewer-level randomization, policy assignment had near-zero effects on paper scores and reviewer confidence. The paper-level randomization also found a near-zero effect on final paper decisions. The same was true for review length: reviews were longer under the permissive policy, although the effect sizes were small. The only outcome that differed across randomizations was review quality: the paper-level randomization found slightly higher review quality ratings under the permissive policy, whereas the reviewer-level randomization yielded a near-zero estimate. Overall, policy assignment had small or near-zero effects across outcomes.\looseness=-1

\subsubsection{Pangram Analysis Results} \label{sec:pangram}

\begin{figure*}[t]
  \centering
  \includegraphics[width=\linewidth]{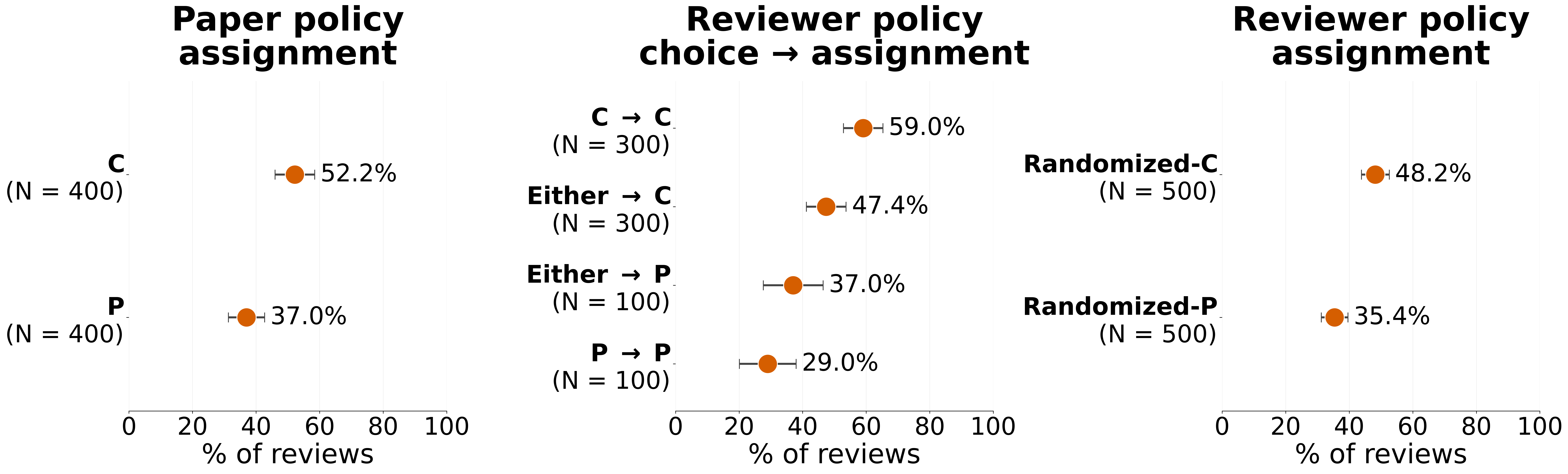}
  \caption{\textbf{Pangram results for a subset of reviews.} Percentage of reviews classified as human-written by Pangram by paper policy assignment (left), reviewer policy choice and assignment (middle), and reviewer policy assignment in the reviewer-level randomization (right). The first two panels are based on reviews sampled from eight strata across all reviews and are reweighted to represent the full review population. The right panel is based on a subset of reviews sampled from the reviews of 940 papers included in the paired-comparison analysis of the reviewer-level randomization. Points show estimated percentages with 95\% confidence intervals. \looseness=-1}
  \label{fig:pangram}
\end{figure*}

Figure~\ref{fig:pangram} shows the percentage of reviews classified by Pangram as human-written rather than AI-generated or mixed. The left panel shows that papers assigned to the conservative policy have higher percentages of human-written reviews: 
C = 52.2\% (95\% CI $[45.9\%, 58.4\%]$), compared to papers assigned to the permissive policy: P = 37.0\% (95\% CI $[31.3\%, 42.7\%]$). (Within the conservative policy, papers that requested and were assigned the conservative policy (C$\rightarrow$C) and papers that permitted the permissive policy but were randomly switched to the conservative policy (P$\rightarrow$C) had similar rates of human-written reviews: 50.6\% (95\% CI $[43.1\%, 58.1\%]$) and 58.8\% (95\% CI $[51.3\%, 66.2\%]$), respectively, with overlapping CIs.)

The middle panel shows that the human-written review percentage is highest among reviewers who chose and were assigned to the conservative policy: C$\rightarrow$C = 59.0\% (95\% CI $[52.8\%, 65.2\%]$). Among reviewers willing to review under either policy, the percentage is higher for those assigned to the conservative policy (Either$\rightarrow$C = 47.4\%, 95\% CI $[41.2\%, 53.7\%]$) than for those assigned to the permissive policy (Either$\rightarrow$P = 37.0\%, 95\% CI $[27.6\%, 46.4\%]$), while it is lowest among reviewers who chose and were assigned to the permissive policy (P$\rightarrow$P = 29.0\%, 95\% CI $[20.1\%, 37.9\%]$). 

Finally, among reviewers randomized to a policy (right panel), those randomized to the conservative policy have a higher percentage of human-written reviews than those randomized to the permissive policy: Randomized-C = 48.2\% (95\% CI $[43.8\%, 52.6\%]$) vs. Randomized-P = 35.4\% (95\% CI $[31.2\%, 39.6\%]$). 

The consistently higher percentage of human-written reviews under the conservative policy suggests that the assigned policy affects reviewers' use of LLMs. At the same time, the results point to substantial noncompliance: even under the conservative policy, only around half of reviews are classified as human-written. While Pangram classifications are imperfect estimates, these results suggest that LLM use may already be widespread in peer review and persists even when its use is prohibited. Indeed, the post-survey also reveals substantial noncompliance (Section~\ref{sec:compliance}). Despite this, these results show that the assigned policy did change reviewers' LLM use, suggesting that the small or near-zero effects of the RCT cannot be explained simply by reviewers ignoring their assigned policy. \looseness=-1

\subsection{RCT Limitations}

There are several considerations for interpreting the RCT findings. First, the results capture the effects of assigning reviewers to different LLM-use policies rather than specific forms of LLM use, as reviewers did not fully comply with their assigned policies. However, compliance rates are also uncertain due to limitations in measurement approaches. Future work should continue efforts to develop more reliable measures of LLM use to better understand the effects of specific forms of LLM use. In addition, our measure of review quality is based on ACs' ratings, leaving open the possibility that LLM-generated reviews may appear higher quality without providing more accurate or useful information, and even less of it. There may also be biases at play, as prior work has found that people sometimes prefer certain forms of AI-generated text to human-written text \cite{porter2024ai}. Future work could complement perceived review quality with additional measures, such as review accuracy, agreement with expert assessments, or the extent to which reviews improve author revisions and editorial decision-making.

\section{Post-Survey on LLM Use, Experiences, and Policy Perspectives}
\label{sec:postsurvey}

\subsection{Post-Survey Methods}

\subsubsection{Survey Design and Data Collection}
To complement the RCT, we conducted an IRB-approved, anonymous post-survey about respondents' LLM use and policy compliance, experiences with LLM-assisted reviewing or reasons for non-use, and perspectives on future LLM-use policies. We describe the questions in more detail as we present the results, and provide the full survey instrument in \cref{app:survey} in the appendix.

Eligible participants included ICML 2026 reviewers, ACs, and SACs for either the main track or the position track (see \cref{method: study context} for role and track descriptions). Note that this differs from the RCT, which was conducted exclusively on the main track. Reviewers were defined as individuals who submitted at least one review for a main-track or position-track submission; ethics reviews did not count toward eligibility. Individuals found to have violated the conference's LLM-use policy by using LLMs when prohibited, approximately 1--2\% of all reviewers, were excluded from participation. \looseness=-1

The survey was administered using Qualtrics and distributed by the PCs to all eligible participants. Before beginning the survey, participants were provided with information about the study and asked to provide informed consent. Participation was voluntary, respondents received no compensation, and responses were collected anonymously. 
Data collection took place over a two-week period, from July 5 to July 19, 2026. During this time, we received 1,486 valid responses, which form the basis of our analysis. Valid responses were defined as survey submissions from individuals who provided consent, indicated that they served as a reviewer, AC, or SAC at ICML 2026, and completed the survey. The median survey completion time was 9.2 minutes (IQR = 5.5--16.1 minutes).

\subsubsection{Data Analysis} 
For close-ended questions, we report both the number of respondents and the corresponding percentage for each item. Because all survey questions were optional, response counts vary across items. Percentages are calculated using the number of respondents who answered the relevant item as the denominator, unless stated otherwise. For open-ended questions, we conducted a thematic analysis. For each question, a member of the research team reviewed a subset of responses and developed an initial codebook, which was subsequently refined through discussion with other members of the research team. The finalized codebook was then applied to all responses with assistance from LLM-based tools and verified by the research team. To support the analysis, we developed a bespoke annotation interface, which is included in Figure \ref{fig:qualitative-coding-interface}. All final decisions regarding the codebook, code assignments, and interpretation of the results were made by the research team.

\subsubsection{Participant Characteristics}

Of the 1,486 respondents, 1,311 (88.2\%) served as reviewers, 160 (10.8\%) as ACs, and 19 (1.3\%) as SACs. Role overlap was rare: only four respondents served as both a reviewer and an AC, and no other overlap was reported. By contrast, many respondents also served as authors, including 940 reviewers (71.7\%), 60 ACs (37.5\%), and 7 SACs (36.8\%). Among reviewers who answered the policy-assignment question (N=1,281), 732 (57.1\%) reported being assigned to the conservative policy and 425 (33.2\%) to the permissive policy. These proportions were broadly consistent with the full reviewer population, in which 63.0\% of reviewers were assigned to the conservative policy and 37.0\% to the permissive policy. The lower percentages reflect the 124 reviewers (9.7\%) who reported being unsure of their assignment. This may be due in part to some reviewers being reluctant to report their policy or to the post-survey being conducted two months after the review period, when some reviewers may no longer remember it. Among reviewers who answered the recognition-award question (N=1,279), 525 (41.0\%) reported receiving a Gold award, 310 (24.2\%) a Silver award, 387 (30.3\%) no award, and 57 (4.5\%) were unsure. Gold reviewers were overrepresented in our sample: in the full reviewer population, 25\% received Gold awards and 25\% received Silver awards. \looseness=-1

Respondents also represented a range of career stages. Among the 1,452 respondents who reported their current position, the largest group was PhD student (37.0\%), followed by Faculty member (32.9\%), Industry researcher (15.1\%), Postdoctoral researcher (13.5\%), Other (0.8\%), Master's student (0.6\%), and Undergraduate student (0.3\%). Regarding PhD timing, 506 of 1414 (35.8\%) had not received a PhD, while 257 (18.2\%) had received a PhD in 2024--2026, 218 (15.4\%) in 2021--2023, and 433 (30.6\%) in 2020 or earlier. Respondents also varied in service experience: among reviewers, the most common level of prior ICML, NeurIPS, or ICLR reviewing experience was 2–3 review cycles (36.6\%, N=1279), while among ACs, the most common level of prior AC experience was 4–6 review cycles (31.8\%, N=157). Finally, 969 of 1450 respondents (66.8\%) reported having spent three or more years at an institution where English was the primary working language. \looseness=-1

\begin{figure*}[t]
  \centering
  \includegraphics[width=\linewidth]{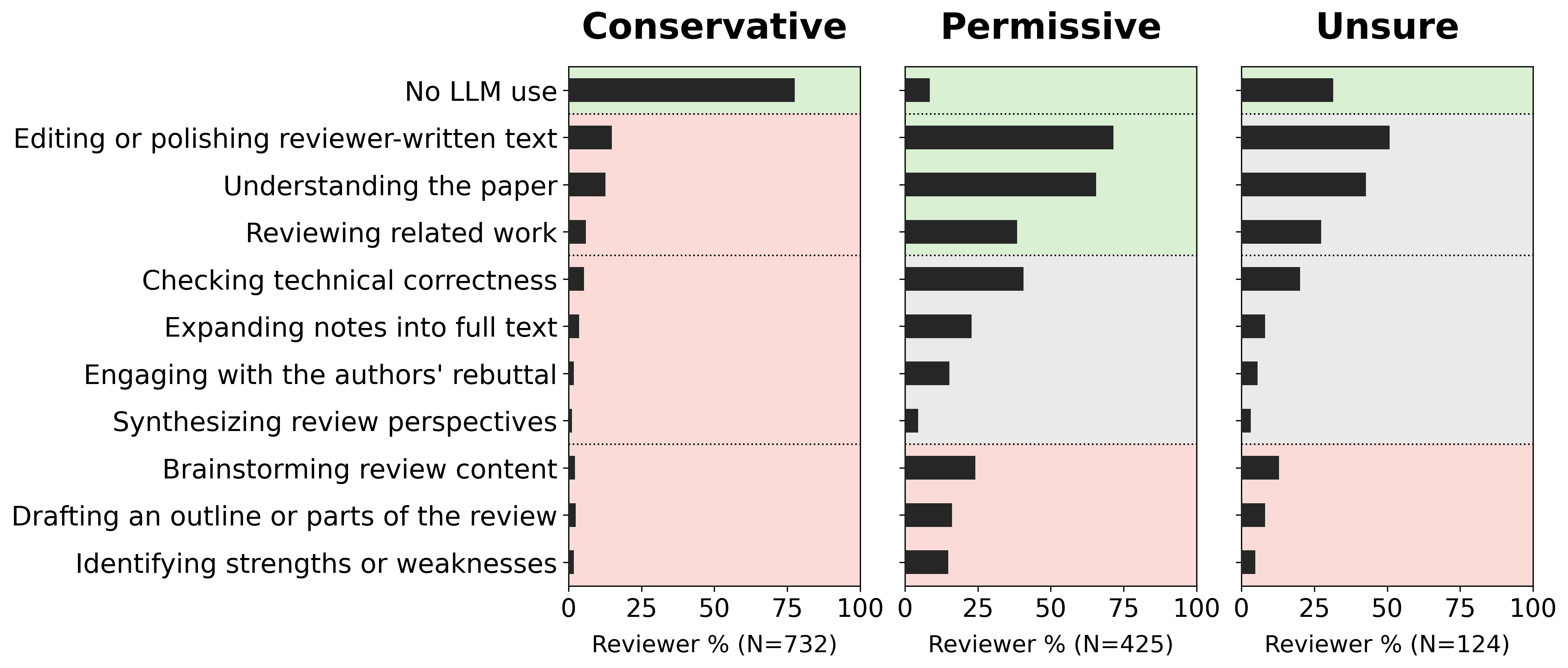}
  \caption{\textbf{Post-survey results showing reviewers’ self-reported LLM use by policy assignment} (N=1,281). Green indicates activities permitted under the assigned policy, red indicates prohibited activities, and gray indicates activities not explicitly covered by the policies. Overall, anonymous respondents frequently reported both prohibited uses and uses not explicitly addressed by the policies. \looseness=-1
  }
  \label{fig:llm-use}
\end{figure*}

\subsection{Post-Survey Results}
\label{sec:postsurveyresults}

We begin by examining reviewers' LLM use and policy compliance (\cref{sec:compliance}), followed by ACs' LLM use (\cref{sec:areachairs}). We then discuss the experiences of reviewers and ACs who used LLMs (\cref{sec:reviewerexperience}) and the perspectives of reviewers and ACs who did not (\cref{sec:nonuse}). Finally, we present perspectives on future LLM policies from reviewers, ACs, and SACs (\cref{sec:communityperspectives}). 

\subsubsection{Reviewers' LLM Use and Policy Compliance}
\label{sec:compliance}

After indicating their role, the first question reviewers answered in the post-survey was whether they had used LLM(s) at any point during the ICML 2026 review process. Of the 1,311 reviewers who answered this question, 662 (50.5\%) reported using an LLM. We next examine their patterns of LLM use and how it aligned with their assigned policy. \looseness=-1

Reviewers described their LLM use through both free-text responses and a multi-select question with the predefined options shown in \cref{fig:llm-use}. Most notably, \textbf{policy noncompliance was common under both policies}: 22.5\% (165 of 732) of the conservative-policy reviewers reported using an LLM despite the prohibition, and 36.5\% (155 of 425) of the permissive-policy reviewers reported at least one explicitly disallowed use. Because Gold reviewers were overrepresented in the survey sample, we reweighted respondents by reviewer recognition status; the resulting noncompliance rates were similar at 23.3\% under the conservative policy and 35.9\% under the permissive policy. These self-reports complement the Pangram analysis in \cref{sec:pangram}.

\textbf{Reported LLM use was concentrated in activities permitted under the permissive policy, but prohibited and gray-area uses were also common.} The most commonly reported LLM uses, across all reviewer groups (conservative policy, permissive policy, and unsure), were editing or polishing reviewer-written text (73.8\%), understanding the paper (66.5\%), and reviewing related work (37.5\%). All three uses were explicitly permitted under the permissive policy, but explicitly prohibited under the conservative policy along with all other LLM use.

However, \textbf{many permissive-policy reviewers also reported prohibited uses}, including brainstorming feedback, questions, or edge cases (24.2\%), drafting an outline or parts of the review (16.2\%), and identifying strengths or weaknesses of the paper (14.8\%). Permissive-policy reviewers likewise reported activities not explicitly addressed by the policy, including checking technical correctness (40.7\%), expanding notes or bullet points into full text (22.8\%), understanding, evaluating, or responding to authors' rebuttals (15.3\%), and synthesizing perspectives across reviews during the discussion phase (4.5\%). Finally, while not reflected in \cref{fig:llm-use}, a small number of respondents described uses not captured by the predefined options, including detecting prompt injections in submissions and critically challenging draft reviews before finalizing them. Together, these findings raise questions about the practical enforceability of activity-based restrictions on LLM use, which we expand in the discussion in \cref{sec:discussion}. \looseness=-1

\textbf{Non-compliant LLM use was often driven by workload pressures and uncertainty about where to draw the line in use.} To better understand why reviewers engaged in non-compliant LLM use, we analyzed free-text responses from 527 reviewers who used LLMs for peer review and answered a post-survey question about whether any of their LLM use may have been inappropriate. Among those who acknowledged inappropriate or gray-area use, common reasons included excessive reviewing workload, time pressure, and uncertainty about the boundary between acceptable LLM assistance and excessive reliance. Respondents frequently mentioned \textit{``the sense of being overwhelmed by review load''} led them to \textit{``rely more heavily on LLM assistance than intended.''} Many also mentioned that it was \textit{``unclear if certain activity is allowed or should be allowed''} when they were using LLMs during review. Respondents also frequently described feeling that some uses became problematic when model outputs influenced their judgments in ways that undermined their sense of ownership over the review. One respondent mentioned that their LLM usage \textit{``may have been slightly inappropriate in the sense that [they] don't feel like [they] completely own [their] review.''} Another respondent said that they may have been sometimes \textit{``influenced by the LLM to change the final judgment.''}

\subsubsection{ACs' LLM Use}
\label{sec:areachairs}

We now turn to ACs, an important yet relatively understudied stakeholder in CS peer review. Although they comprise a much smaller population than reviewers, ACs play a central role in the review process by recruiting and managing reviewers, monitoring review quality, facilitating discussions, preparing meta-reviews, and making recommendations on submitted papers. At ICML 2026, ACs did not receive specific guidance on LLM use, unlike reviewers. Among the 160 respondents who served as an AC, 75 (46.9\%) reported using an LLM to support their AC duties, while 85 (53.1\%) did not. This adoption rate is comparable to that of reviewers (50.5\%), despite the different responsibilities and policy guidance associated with the two roles. \looseness=-1

To better understand how ACs used LLMs, we coded free-text responses from the 72 ACs who described their use. Because ACs' use of LLMs remains relatively understudied, we used an open-ended question rather than predefined activities to capture a broader range of uses. The most commonly reported uses were editing or polishing writing (47.2\%) and synthesizing perspectives across reviews during the discussion phase (45.8\%), reflecting ACs' role in consolidating reviewer feedback into meta-reviews. Smaller but notable shares of ACs reported using LLMs to understand papers (16.7\%), check meta-reviews for completeness or accuracy (13.9\%), and engage with authors' rebuttals (12.5\%). Less common uses included expanding notes into full text (8.3\%), identifying strengths and weaknesses of papers (8.3\%), drafting portions of meta-reviews (5.6\%), reviewing related work (2.8\%), and brainstorming content for meta-reviews (1.4\%). Compared with reviewers, whose most common use was also editing or polishing reviewer-written text but whose other top uses included understanding papers and reviewing related work, AC use was more concentrated on synthesis and meta-review preparation tasks. \looseness=-1

\begin{figure*}[t] 
\centering 
\includegraphics[width=\linewidth]{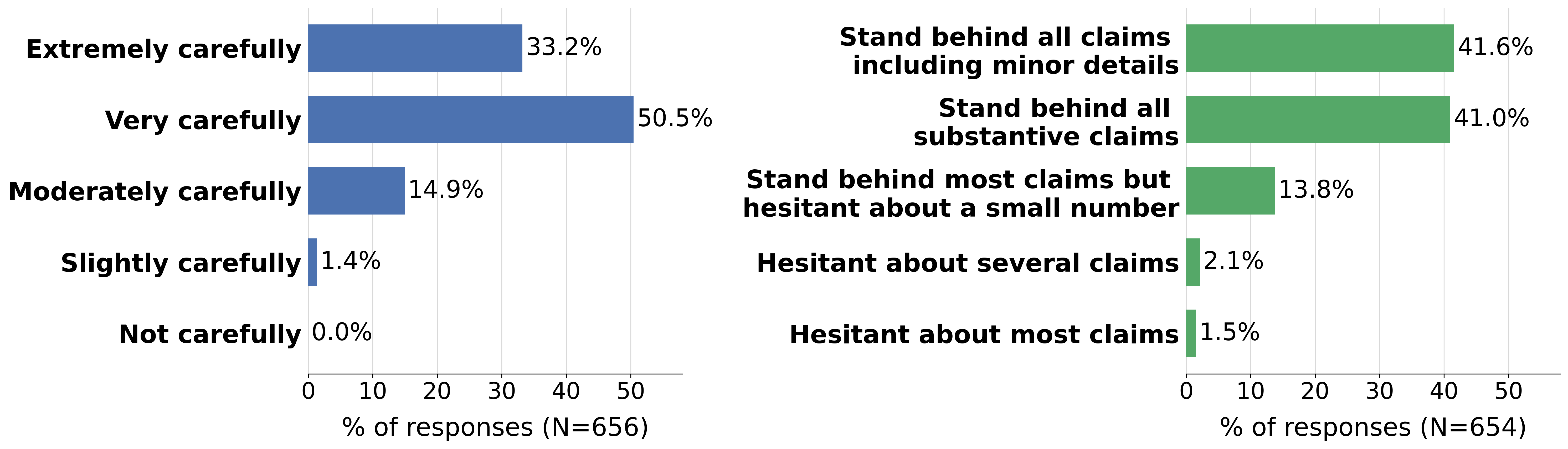} 
\caption{\textbf{Post-survey results on how carefully LLM-using reviewers reported reading and verifying LLM outputs (left) and the extent to which they stand behind the claims in their reviews (right)}. Most reviewers reported carefully reviewing LLM outputs and largely standing behind the claims in their reviews. However, we emphasize that these results are based on self-reports rather than direct observations of how reviewers interacted with LLMs.}
\label{fig:careful_standby}
\includegraphics[width=\linewidth]{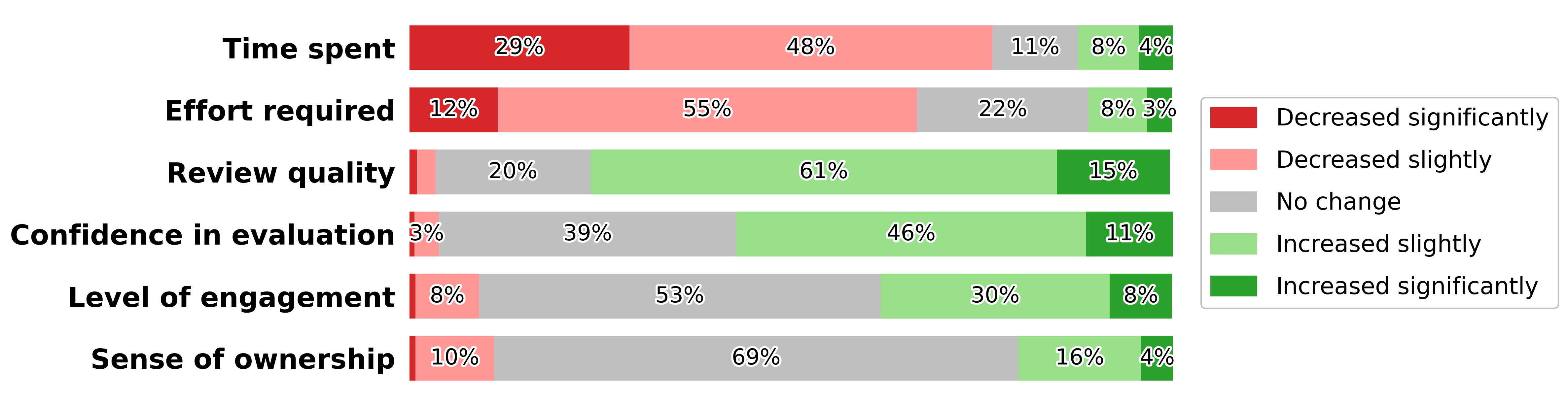} 
\caption{\textbf{Post-survey results on reviewers' self-reported effects of LLM use} (N=647–650 per item). Reviewers who used LLMs generally reported positive effects, particularly improved review quality and reduced time spent, although experiences varied across respondents. Non-users did not answer these questions; however, elsewhere in the survey, many reported little need for LLM assistance or expressed principled objections to its use (\cref{sec:nonuse}).
}
\label{fig:reviewer-perceptions}
\end{figure*}

\subsubsection{Reviewer and AC Experiences with LLMs}
\label{sec:reviewerexperience}

We next examine the experiences of reviewers and ACs who used LLMs. Reviewers who reported using LLMs were asked how carefully they read and verified LLM outputs, the extent to which they stood behind the claims in their reviews, and how LLM use affected various aspects of reviewing. ACs who reported using LLMs were asked analogous questions about the effects of LLM use on their AC duties. For reviewers, we also conducted exploratory analyses by reviewer recognition status, current position, PhD status and timing, review experience, and English-language experience. We report only the most notable differences observed; responses were otherwise broadly similar across the reviewer characteristics examined.

\textbf{Most reviewers who used LLMs reported carefully verifying LLM outputs and standing behind the claims in their reviews}. As shown in \cref{fig:careful_standby}, 83.7\% of 656 respondents selected ``extremely carefully'' or ``very carefully'' reading and verifying LLM outputs, and 82.6\% of 654 respondents reported standing behind all substantive claims in their reviews, including 41.6\% who reported standing behind all claims, including minor details. 
Among the reviewer characteristics examined, the largest differences were observed by reviewer recognition status and English-language experience. Gold reviewers and reviewers with more English-language experience were more likely to report reading and verifying LLM outputs ``extremely carefully" and standing behind all claims in their reviews, including minor details. \looseness=-1

\textbf{Reviewers who used LLMs generally reported positive effects, most notably improved review quality and reduced time spent, but experiences varied across respondents.} As shown in \cref{fig:reviewer-perceptions}, among the 647--650 LLM-using reviewers who answered each item, 76.3\% reported reduced time spent and 66.6\% reported reduced effort. Similarly, 76.2\% reported improved review quality and 57.2\% reported increased confidence in their evaluations. At the same time, nontrivial minorities reported the opposite pattern: 12.5\% reported increased time spent, 10.9\% reported increased effort, 3.4\% reported reduced review quality, and 3.8\% reported reduced confidence. Responses were more mixed for engagement and sense of ownership, for which no change was the most common response (52.7\% and 68.6\%, respectively), though respondents also reported both increases (38.2\% and 20.3\%) and decreases (9.1\% and 11.1\%). Among the reviewer characteristics examined, the largest differences were again observed by reviewer recognition status: Gold and Silver reviewers reported larger increases in confidence in their evaluations than reviewers who received no award, while Silver reviewers reported larger reductions in time spent.

\textbf{ACs who used LLMs reported patterns broadly similar to those observed among reviewers.} Among the 73 LLM-using ACs who answered analogous questions, 72.6\% reported reduced time spent, 74.0\% reported reduced effort, 68.5\% reported improved meta-review quality, and 50.7\% reported increased confidence in their evaluations, broadly mirroring the corresponding reviewer results. Responses were more mixed for engagement in AC duties, sense of ownership over the meta-review, and depth of interaction with reviewers: no change was the most common response (64.4\%, 65.8\%, and 74.0\%, respectively), though many respondents reported both increases (31.5\%, 20.5\%, and 17.8\%, respectively) and decreases (4.1\%, 13.7\%, and 8.2\%, respectively). See \cref{fig:ac-perceptions} for the full distribution of responses. \looseness=-1

\subsubsection{Reviewer and AC Reasons for Not Using LLMs}
\label{sec:nonuse}

We now shift our attention to the 649 (49.5\%) reviewers and 85 (53.1\%) ACs who reported not using an LLM. Of these respondents, 609 reviewers and 82 ACs provided a free-text explanation of their decision.  Among reviewers, \textbf{policy compliance} was the most commonly cited reason (61.7\%), although many of these respondents also cited other concerns. ACs also cited policy-related reasons (30.5\%), such as uncertainty about whether LLM use was permitted, likely because they were not given specific guidance on LLM use. Some respondents in both groups also indicated that \textbf{LLMs are not needed} (12.2\% of reviewers; 30.5\% of ACs), citing their expertise, established workflows (e.g., \textit{``I have my process for reviewing papers. It works nicely, and it's routine''}), or skepticism about the benefits of LLM use (e.g., \textit{``verifying LLM works seems about as time consuming as just doing the AC work without it''}). \looseness=-1

\textbf{Distrust in LLMs' ability to perform review-related tasks} was a common concern among both reviewers (21.5\%) and ACs (23.2\%). Respondents described LLM-generated evaluations as generic, lacking nuance, and prone to hallucinations. As one reviewer put it, LLMs' \textit{``critiques are shallow, unfair, and lack balanced perspectives.''} Closely related were \textbf{concerns about preserving independent judgment, ownership, and responsibility} (22.2\% of reviewers; 23.2\% of ACs). Respondents emphasized the importance of reaching their own conclusions rather than allowing LLM output to \textit{``bias''} their evaluations. One reviewer stated, \textit{``I wanted the review to reflect my own understanding and independent judgment of the submission.''} Others worried that relying on LLMs could lead to a loss of \textit{``cognitive ownership''} of the review or constitute an \textit{``abdication of responsibility.''} \looseness=-1

The view that \textit{``peer review is the responsibility of peers, and not machines''} was shared by 11.8\% of reviewers and 22.0\% of ACs, who opposed using LLMs for peer review regardless of their capabilities. They described that \textbf{peer review is an inherently human activity}. As one reviewer put it: \textit{``We should not be giving our human role as scientists to LLMs. If we do, then it will ruin the profession.''} Others emphasized the \textbf{responsibility to authors}, describing LLM-assisted review as \textit{``unethical,''} \textit{``disrespectful,''} or not the kind of review they would want for their own work. One reviewer wrote, \textit{``I would like to think that the people who review my papers have at least read the paper by themselves and formed their own opinion about it. I feel that it is only fair that I should do the same for other people's papers.''}

A small number of respondents also raised \textbf{privacy and confidentiality concerns}. For example, one reviewer wrote: \textit{``I prefer not to violate the privacy and confidentiality of author's intellectual property by sharing their work with proprietary third party software.''} Some cautioned that widespread LLM use could \textbf{compromise the diversity of perspectives} in peer review, emphasizing that \textit{``human reviewers come from different backgrounds and experiences which bring unique viewpoints and experiences to the table during the review process.''} Others worried that it could \textbf{incentivize writing for LLMs rather than humans}, with one respondent warning of a future \textit{``where LLMs are used to write papers that pass LLM reviews.''} Finally, some respondents raised \textbf{environmental concerns} about the resource demands of LLMs. \looseness=-1

\begin{figure*}[t]
  \centering
  \includegraphics[width=\linewidth]{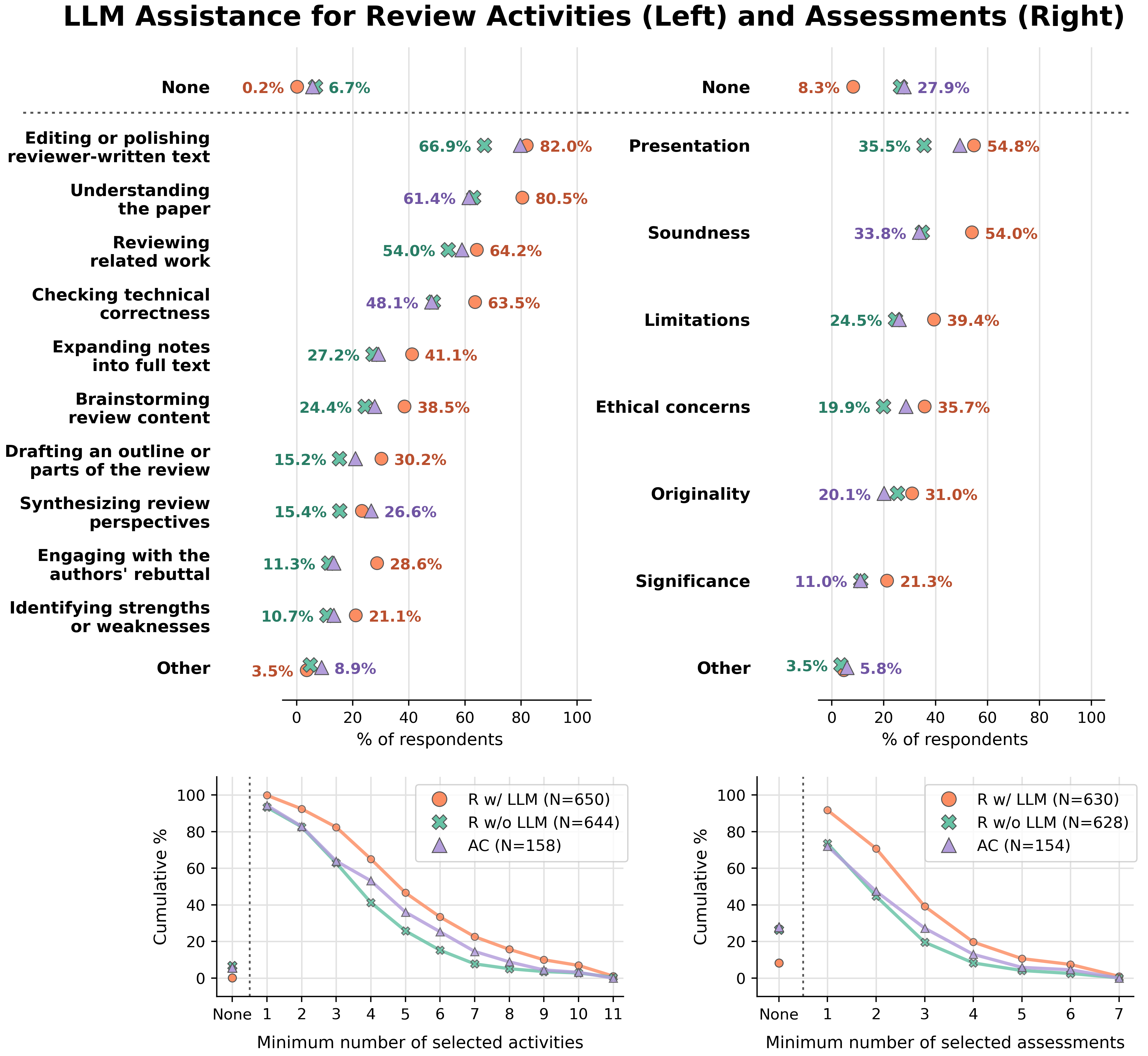}
  \caption{\textbf{Post-survey results on which activities reviewers should be allowed to use LLMs for (left) and which aspects of a paper should be assessable with LLMs (right), assuming full policy compliance.} The top panels show the percentage selecting each option among reviewers who reported using LLMs (circles), reviewers who reported not using LLMs (Xs), and ACs (triangles), with only the lowest and highest percentages labeled. The bottom panels show the cumulative percentages selecting at least a given number of activities or aspects. Reviewers who reported using LLMs generally supported a broader range of LLM uses than ACs or reviewers who reported not using LLMs.}
  \label{fig:policy-preferences}
\end{figure*}

\subsubsection{Community Perspectives on Future LLM Policies}
\label{sec:communityperspectives}

In the final part of the post-survey, all respondents (reviewers, ACs, and SACs) were asked about their views on future policies governing reviewer LLM use. Overall, reviewers, ACs, and SACs were largely aligned in their views. We note where this was not the case.

\textbf{Respondents broadly supported LLM use for assistive tasks but were more cautious about tasks requiring substantial reviewer judgment.} To understand which forms of LLM use the community considers acceptable, we asked respondents which review activities and assessment of which paper aspects (e.g., presentation, soundness, originality) LLM assistance should be allowed for, \textit{assuming full reviewer compliance}. As shown in \cref{fig:policy-preferences}, respondents across all groups (reviewers who reported using LLMs, reviewers who reported not using LLMs, and ACs) were most supportive of using LLMs to edit reviewer-written text, understand the paper, or review related work, the three use cases explicitly allowed under ICML 2026's permissive policy. In contrast, they expressed substantially less support for using LLMs to draft an outline or parts of a review, identify the paper's strengths and weaknesses, or assess its originality and significance. \looseness=-1

Overall, only a small fraction selected ``None'' for review activities (0.2--6.7\% across groups), indicating that at least 93.3\% of each group supported LLM use for at least one activity. This included many respondents who did not use LLMs themselves and had expressed substantial concerns about LLM use in \cref{sec:nonuse}. Selecting ``None'' was more common for the question of which paper aspects reviewers should be allowed to use LLMs to assess (8.3--27.9\%), although at least 72.1\% of each group still supported LLM use for at least one aspect. While the groups were largely aligned, reviewers who reported using LLMs supported a broader range of LLM uses than both ACs and reviewers who reported not using LLMs. Another noteworthy observation is that compared with reviewers, ACs were less supportive of using LLMs to understand papers or check technical correctness, but more supportive of using them to synthesize perspectives across reviews, which more closely reflects their tasks. \looseness=-1

Free-text responses on inappropriate LLM use in peer review provided further insights into respondents' views on acceptable LLM assistance. In open-ended questions, respondents described cases from their own experience or from instances they had observed or heard about. \textbf{Respondents generally viewed appropriate use as restricting LLMs to supportive roles and retaining independent judgment and accountability for their reviews.} In contrast, uses described as inappropriate included submitting predominantly LLM-generated reviews, submitting reviews containing hallucinated references or critiques, and using LLMs to justify predetermined paper decision recommendations. \looseness=-1

\textbf{For many respondents, policy enforceability was the central concern.} We next asked respondents whether their preferred policy would change if reviewers did not fully comply with whatever policy was adopted. Among the 723 respondents who provided a free-text response, 315 (43.6\%) explicitly stated that their preferred policy would remain unchanged, while only 29 (4.0\%) said it would change. The remaining 379 (52.4\%) respondents described their preferred policy without explicitly indicating whether it differed from their earlier answer.

Across responses, the dominant theme was that enforcing any policy on LLM use would be extremely difficult. Respondents frequently argued that reviewers cannot realistically be prevented from feeding papers to LLMs and that distinguishing AI-polished from AI-generated text is unreliable. Several respondents commended ICML 2026's watermarking mechanism for detecting policy violations (\cref{sec:policyassignment}), but argued that it was \textit{``already obsolete''} because both reviewers and newer models could route around it. As a result, many proposed that \textbf{enforcement should focus on holding reviewers accountable for the reviews they sign their name to, rather than policing LLM use}. Suggested consequences for low-quality or clearly LLM-generated reviews included \textit{``desk rejection of [reviewers'] own papers''} and \textit{``multi-year reviewing/submission bans.''} Another recurring proposal was that \textbf{conferences should provide a vetted, logged LLM} so that LLM use can be monitored directly and made available to all reviewers.

Among the small minority who explicitly stated in the survey that they would change their preferred policy when the full-compliance assumption was relaxed, opinions diverged in opposite directions but shared the same underlying concern: enforceability. Some argued that if activity-based restrictions cannot be reliably enforced, any permitted use of LLMs will gradually expand in practice, making \textbf{a complete ban on LLM use} the only defensible option even if occasional violations remain inevitable. Respondents described permitted uses as a \textit{``slippery slope''} and argued that \textit{``any permitted use of LLMs will mentally encourage being lazy and doing the full review with an LLM.''} Others, however, argued that unenforceable restrictions mainly \textit{``punish honest reviewers while doing nothing about bad actors,''} making it preferable to \textnormal{permit broad LLM use and focus oversight on review quality rather than the methods used to produce it}.

\textbf{Support for piloting fully automated, clearly labeled LLM-generated reviews alongside human reviews was mixed.} Inspired by AAAI's 2026 pilot~\cite{aaai2026aireview,biswas2026aaai}, we asked respondents whether ICML should consider piloting such reviews as an additional input during the initial review stage, with authors and ACs explicitly informed that the reviews were LLM-generated. Among 1,457 respondents, 786 (53.9\%) supported some form of pilot: 462 (31.7\%) supported an optional, opt-in service for authors, and 324 (22.2\%) supported providing LLM-generated reviews for all submissions. In contrast, 436 (29.9\%) indicated that ICML should not use fully automated reviews, while 235 (16.1\%) were unsure. Overall, views were broadly aligned across reviewers, ACs, and SACs.

Finally, we asked about practices that could help reviewers produce high-quality reviews while maintaining strong ownership over their reviews. 826 respondents answered this question, and their recommendations fell into two categories: individual responsibility and institutional support. First, respondents emphasized that \textbf{reviewers should restrict LLMs to assistive tasks and reserve evaluative judgments for themselves}. These tasks included understanding the paper, identifying overlooked related work, and polishing the review. As one respondent put it: \textit{``the key is to separate judgment from expression: form all evaluations independently first, then use the LLM only to refine the wording of your own comments, never to generate opinions or process the manuscript itself.''} Respondents also emphasized that \textbf{reviewers should verify all LLM outputs against the manuscript before incorporating them into their reviews}. This restriction was paired with a verification obligation almost without exception: nothing an LLM produces (an interpretation, a citation, a rephrased sentence) should enter the review unchecked. As one respondent put it, reviewers should \textit{``verify every edited sentence still matches their intended meaning, avoid pasting confidential content into any model, and disclose their use.''}
\looseness=-1

A second cluster of recommendations focused on how conferences can support responsible LLM use, arguing that individual responsibility alone is insufficient.
Consistent with respondents' explanations for policy noncompliance (\cref{sec:compliance}), the most frequently cited structural lever was reviewing workload. Respondents recommended that \textbf{conferences should reduce papers assigned per reviewer and improve reviewer-paper matching}, arguing that \textit{``the sense of being overwhelmed by review load is probably what drives many reviewers to LLMs in the first place,"} and that reviewers matched to papers within their expertise have less reason to \textit{``lean on [an] LLM."} Respondents also argued that \textbf{high-quality reviewing should be better rewarded} through recognition, registration waivers, or reputational credit, making reviewing feel \textit{``valuable and rewarded rather than like an unpaid tax on doing research."} Additional recommendations included that \textbf{conferences should provide tools and clear guidance on appropriate LLM use}. Respondents advocated for a \textit{``conference-hosted, privacy-preserving LLM restricted to approved tasks,''} as well as a \textit{``curated set of allowed prompts and concrete published guidance,''} including worked examples and explicit do's and don'ts, so that appropriate use is not left to individual interpretation.

\subsection{Post-Survey Limitations}

We close this section by reflecting on a few limitations of the post-survey. First, it may be subject to selection and response bias. Although the survey received responses from 1,486 reviewers, ACs, and SACs, this represents only a small fraction of the full program committee (17,886 reviewers, 1,728 ACs, and 251 SACs). Gold reviewers were also overrepresented among respondents (41.0\% vs. 25\% in the full reviewer population), suggesting that our respondents may reflect a more engaged subset of the broader reviewer population.

Next, self-reported LLM use and perceived effects may not fully capture actual use or effects. Prior work suggests that people can misperceive their AI use and overestimate its benefits~\cite{yu2026efficiencygainillusion,Fernandes2026}. Similarly, in an experimental study of CS peer review, \citet[Section 10.1.2]{shah2022acm} found no correlation between reviewers' self-reported confidence or expertise and review quality, as assessed by objective criteria such as the absence of false or unsubstantiated claims and the presence of specific feedback. Thus, future work should examine whether these perceived benefits translate into improvements in review quality and reviewer experiences. One direction would be to combine survey responses with behavioral measures and review outcomes at the participant level. In particular, conference-provided LLM tools could provide a richer understanding of how reviewers use LLMs by capturing prompts, outputs, and usage patterns alongside review outcomes. \looseness=-1

Finally, our findings from both the RCT and post-survey are situated within a single conference and review cycle, while both LLM capabilities and norms governing their use in peer review are evolving rapidly. Future work could examine how these changes shape reviewer behavior and the effectiveness of peer-review systems over time. It could also investigate the perspectives of stakeholders beyond reviewers, ACs, and SACs, such as readers, academic departments, and other consumers of peer-review signals. These groups may rely on peer review primarily for quality control and curation rather than participation in the review process itself, and may have different interests and concerns regarding LLM use. \looseness=-1

\section{Discussion}
\label{sec:discussion}

\subsection{Implications for the Design of Policies and Tools for Peer Review}
\label{sec:disc_policy}

In this section, we reflect on our findings and discuss promising directions for future peer-review policies and tools.

\subsubsection{Design Realistic Policies and Cultivate the Right Environment} \label{sec:realistic-policy}

Traditionally, reviewing has been a valued, voluntary activity. In a survey of 307 CHI reviewers, \citet{nobarany2016motivates} found that ``encouraging high-quality research, giving back to the research community, and finding out about new research'' were reviewers' primary motivations for reviewing. Our post-survey responses echoed these themes, describing science as a \textit{``community process''} and emphasizing that \textit{``the fundamental purpose of peer review [...] is to provide expert, accountable, and independent judgment.''} These findings suggest that, with the right culture and environment in place, reviewers would be motivated to conduct high-quality reviews. \looseness=-1

However, our findings suggest that these conditions are not in place, as evidenced by the high rates of noncompliance in both the Pangram analysis (\cref{sec:pangram}) and post-survey results (\cref{sec:compliance}).
These findings also suggest that activity-based LLM policies are difficult to enforce in practice, particularly when enforcement relies on post-hoc detection of how reviewers used external tools. In the absence of robust enforcement mechanisms, conferences may be better served by \textbf{cultivating norms, providing guidance, and aligning incentives} rather than imposing strict but unenforceable restrictions. 
For example, many post-survey respondents reported uncertainty about where the boundaries of appropriate LLM use lie (\cref{sec:nonuse}). 
Rather than relying solely on high-level policy statements, conferences could provide concrete guidance and illustrative examples that help reviewers distinguish between acceptable and prohibited uses. We expand what this guidance could look like in \cref{sec:judgement-ownership}.

Furthermore, future conferences could create a culture in which reviewers feel comfortable 
disclosing a broader range of LLM uses beyond simply ``polishing the review,'' without stigmatizing such disclosures. Indeed, our post-survey respondents generally viewed several additional forms of LLM assistance as permissible, including using LLMs to understand the submission or identify potential technical errors (\cref{sec:communityperspectives}). In line with what has been proposed for paper writing \cite{feuerriegel2026reporting}, one possible approach could be to introduce an ``LLM usage form'' that allows reviewers to document how they used LLMs during the review process. For example, a reviewer might report: {``I used an LLM with multiple prompts and follow-up questions to help verify whether Theorem 2 was correct.''} Such disclosures could promote transparency while providing the research community with greater visibility into emerging review practices. \looseness=-1

Finally, our research represents an initial step toward generating empirical evidence to inform peer-review policy design. In our post-survey, respondents positively commented on the opportunity to share their thoughts and influence future policies. As LLM capabilities and reviewing practices continue to evolve, ongoing input and deliberation from reviewers, authors, conference organizers, the broader research community, and the public will be essential to keeping policies effective and realistic.

\subsubsection{Manage Submission Volume and Incentivize Reviewers} \label{sec:peer-review-system}

In our post-survey, many respondents cited excessive workload and time pressure as reasons for relying on LLMs in ways that stretched policy boundaries (\cref{sec:compliance}), suggesting that realistic LLM policies must account for these pressures. However, these challenges predate LLMs: peer review systems have long struggled with growing submission volumes and review workloads \cite{beygelzimer2023neurips,nobarany2016motivates,shah2022acm}, and researchers continue to explore ways to address them~\cite{icml2026deskreject, iclr2025feedbackagent, biswas2026aaai, neurips2026experiment, emnlp2026experiment}.
Building on these efforts, we synthesize existing and emerging approaches to highlight promising directions for reducing the pressures that may drive LLM use. \looseness=-1

The first direction is to continue strengthening efforts to \textbf{manage submission volume}, including authorship quotas~\cite{TMLRquotas2026,shah2026many}, desk rejections~\cite{NeurIPS2020deskrejections}, and rethinking metrics and criteria for paper acceptances in the LLM era. Increasing numbers of submissions may be geared toward inflating publication records, with authors providing little oversight of AI-generated manuscripts \cite{gillespie2025trust, spinellis2025false, palivela2026stop}. Reviewing such work can erode reviewers' motivation, as their feedback is unlikely to support genuine learning or research improvement. Strengthening controls on what enters the review pipeline could therefore help reduce review burden and preserve a culture in which reviewers feel responsible for supporting fellow researchers.

At the same time, conferences should also \textbf{provide sufficient incentives} for peer review. For example, ICML 2026 gave Gold awards and free conference registration to top reviewers, while HCI conferences such as CHI and CSCW offer review recognitions awarded by ACs. However, financial incentives of this kind may not be sustainable or feasible for every conference. Conference should therefore continue exploring ways to recognize high-quality reviewers and reward thoughtful reviews through a range of \textit{external} incentives. There may also be opportunities to foster \textit{internal} motivation for high-quality reviewing, which we discuss in \cref{sec:preserving-by-product}.

\subsubsection{Explore Conference-Provided Reviewing Tools} \label{sec:conference-llm}

Beyond policy, our findings point to opportunities for tool design and experimentation. As currently being explored by NeurIPS 2026 \cite{neurips2026experiment}, conference-provided LLM tools could standardize access, privacy protections, and permitted forms of use. 
Conferences could also provide supporting resources, such as exemplar prompts and skill files tailored to peer-review tasks, as suggested by many respondents (\cref{sec:communityperspectives}).
These vetted resources could address several needs raised by respondents (\cref{sec:reviewerexperience}), including clearer guidance on acceptable LLM use, stronger privacy protections for unpublished manuscripts, and more equitable access to reviewing assistance. With appropriate reviewer consent, they could also generate valuable data on how LLMs are used in practice, informing future policy and tool design. Future work could also draw inspiration from existing HCI and CSCW systems designed to support academic workflows \cite{sun2024reviewflow}. 
More generally, we encourage viewing conference-provided tools not only as a means of supporting compliance, since providing such tools alone would not ensure that reviewers use only approved tools, but also as an opportunity to explore broader sociotechnical approaches to AI governance and reviewer support.

\subsection{Supporting Meaningful Human Engagement in Peer Review and Beyond}
\label{sec:disc_engagement}
 
As discussed in \cref{rw:human}, peer review represents one form of consequential knowledge work in which people make judgments that affect others. Beyond academic reviewing, our findings speak to a central question in HCI: how can AI systems provide useful assistance while preserving human judgment, expertise, ownership, and accountability?

\subsubsection{Reflection on Human Judgment and Ownership in Peer Review} \label{sec:judgement-ownership}

Our findings extend prior HCI work on human judgment and ownership in human-AI interaction~\cite{Jakesch2023Cowriting, anderson2024homogenization, padmakumar2024does,doshi2024generative,ashkinaze2025how, kim2024uncertainty,kim2025fostering,ibrahim2026aies} 
by identifying which aspects of peer review the academic community sees as appropriate for LLM assistance and which they see as requiring independent human judgment. In the post-survey, respondents broadly supported LLM use for assistive tasks such as understanding submissions, reviewing related work, and improving writing, but were more cautious about core evaluative tasks such as identifying strengths and weaknesses of submissions, assessing their originality and significance, and drafting reviews (\cref{fig:policy-preferences}). These preferences also reflect respondents' emphasis on preserving independent judgment, ownership, and responsibility in peer review (\cref{sec:nonuse}). \looseness=-1

Our work also offers a more nuanced perspective on ownership in AI-assisted work. In the post-survey, reviewers who used LLMs reported mixed effects on their sense of ownership over their reviews: 68.6\% reported no change, while 20.3\% reported increased ownership and 11.1\% reported decreased ownership (\cref{fig:reviewer-perceptions}). We do not know what drove these differences, but respondents' comments offer clues about how they understood ownership. In particular, several respondents emphasized \textit{provenance} (e.g., \textit{``the critical analysis, scientific judgment, and final evaluation must always come from the reviewer, not from the LLM''}) and \textit{defensibility} (e.g., \textit{``if the reviewer can personally defend every claim in the review, ownership is preserved'')}. Our post-survey included questions about reviewers' sense of ownership and how much they stood behind the claims in their reviews (\textit{defensibility}). Future work could additionally assess \textit{provenance}, for example, by measuring how much of the review content originated from the reviewer versus an LLM. More broadly, these findings highlight the need to better understand and measure ownership in AI-assisted knowledge work. \looseness=-1

\subsubsection{Preserving the Values and Byproducts of Knowledge Work} \label{sec:preserving-by-product}

To sustain a healthy peer review ecosystem, the goal is not only to produce high-quality reviews, but also to preserve the broader values and byproducts of reviewing for reviewers: learning new topics, growing as a researcher, and fostering a sense of community within a field \cite{nobarany2016motivates, sun2024reviewflow}. 
This view was also shared by our respondents, who described \textit{``learning more about the background of [a] particular subfield''} and \textit{``understanding how to write better review''} as important goals of conducting peer review. One way to preserve these values is to focus on the byproducts of knowledge work and design tools to support them, an area where HCI researchers have deep expertise and are well positioned to contribute.

For example, research could focus on building systems that serve as thought partners rather than text editors and generators, supporting reviewers' understanding and reasoning. Consistent with the best practices identified by respondents (Section~\ref{sec:communityperspectives}), such systems could help reviewers articulate and refine their own observations, connect concerns to evidence, and understand how specific issues influence evaluation decisions. More broadly, writing assistance should be grounded in reviewers' evolving reasoning and support their development as reviewers. 

A second direction could be to support reviewers' learning from one another. Building on social computing research in HCI and CSCW \cite{dow2012shepherding, morris2007searchtogether, geiger2014personalized, cosley2007suggestbot, deng2025weaudit}, future systems could help facilitate communication between reviewers and ACs, surface unresolved disagreements and their underlying rationales, and help reviewers learn how other experts interpret evidence and reach decisions. Such systems could create opportunities for reviewers to engage with and learn from others, preserving some of the social value of participating in a shared review community. 

Taken together, we hope these ideas could inspire the community to look beyond ``fixing" or governing LLM use in peer review, and consider new ways to support its broader value. In particular, we encourage continued research and engagement between conference organizers and the research community. Our work represents one step toward that goal. \looseness=-1

\section*{Acknowledgments} 
The work of NBS was supported by NSF grant 1942124.


\bibliographystyle{ACM-Reference-Format}
\bibliography{references}

\clearpage
\appendix
\section*{Appendix}

The appendix is organized as follows:
\begin{itemize}
    \item \cref{sec:llmpolicies}: LLM Policies in Computer Science Peer Review
    \item \cref{sec:postsurveyadd}: LLM Tools Used by Reviewers and ACs
    \item \cref{app:survey}: Full Survey Instrument
    \item \cref{sec:reviewform}: ICML 2026 Main-Track Review Form
    \item \cref{sec:acreviewquality}: ICML 2026 AC Review Quality Rating

\end{itemize}

\section{LLM Policies in Computer Science Peer Review}
\label{sec:llmpolicies}

In this section, we review the rapidly evolving landscape of LLM policies for peer review at computer science conferences to further contextualize our work. Overall, ICML 2026's two policies are broadly consistent with those adopted by other conferences (see \cref{method: study context} for details). What distinguishes ICML 2026 is its dual-policy framework, which accommodates author and reviewer preferences, and its use of a controlled experiment to examine the effects of the two policies.

In response to the growing use of LLMs in peer review, computer science conferences have begun developing policies governing when and how reviewers may use these tools. To better understand this evolving policy landscape, we surveyed peer-review LLM policies at HCI, ML, computer vision (CV), natural language processing (NLP), and general-science venues over the 2025--2026 review cycles. Although policies vary, all emphasize reviewers forming independent judgments, remaining accountable for their reviews, and restricting the sharing of paper or review content with tools that may compromise confidentiality. Within these shared constraints, we identify four broad policy categories. \looseness=-1

First, some venues prohibit reviewer LLM use altogether (e.g., ICML 2025 \cite{icml2025policy}, AISTATS 2026 \cite{aistats2025policy}, CVPR 2026 \cite{cvpr2026policy}, ECCV 2026 \cite{eccv2026policy}, and ICCV 2025 \cite{iccv2025policy}). Second, many venues restrict LLM use to language-level assistance, such as grammar, spelling, or clarity edits (e.g., AISTATS 2025 \cite{aistats2025policy}, ACM venues such as CHI, CSCW, and FAccT, as well as Nature \cite{naturepolicy}and PNAS \cite{pnaspolicy}). Third, some venues additionally permit conceptual or background assistance (e.g., NeurIPS 2025 \cite{neurips2025policy}, TMLR \cite{TMLRaireviews2026}, and ACL Rolling Review\cite{arrpolicy}). For example, NeurIPS permits using LLMs to improve understanding of concepts \cite{neurips2025policy}, while ACL Rolling Review permits tools that help check proofs \cite{arrpolicy}. Finally, a smaller number of venues do not explicitly prohibit particular forms of LLM assistance, instead emphasizing disclosure and reviewer accountability (e.g., ICLR 2026 \cite{iclr2026policy} and COLM 2026 \cite{colm2026policy}).

Despite these broad categories, individual policies vary in important ways. In particular, venues differ in whether paper or review content may be entered into an LLM and whether LLM use must be disclosed. For example, within the second category, ACM venues explicitly prohibit sharing paper content with LLMs, whereas Nature permits it provided the content is not shared with ``unsecured or public AI tools.''
Disclosure requirements also vary: for instance, ICLR 2026 and COLM 2026 require reviewers to disclose LLM use ``whether cosmetic or not," whereas many other venues do not, including ACM venues that permit more limited forms of LLM assistance. 

To keep pace with shifting norms and emerging LLM capabilities, CS conference policies are rapidly adapting by piloting new approaches and conducting experiments to explore the role of LLMs in peer review. For example, ICLR 2025 piloted an AI agent that provided optional feedback on reviewers' existing reviews, with reviewers deciding whether to incorporate the suggestions \cite{iclr2025feedbackagent}. AAAI 2026 introduced AI-generated reviews and discussion summaries to supplement human reviews, while leaving evaluation and decision making to human reviewers and program committee members \cite{aaai2026aireview}. Concurrent with this work, NeurIPS 2026 is conducting an opt-in randomized experiment in which participating reviewers are assigned to unassisted, open-ended LLM-assisted, or structured LLM-assisted reviewing conditions \cite{neurips2026experiment}. TMLR has evaluated multiple AI reviewers for assessing soundness of submitted papers, and is deploying one of them~\cite{TMLRaireviews2026,rao2026evaluating}. EMNLP 2026 is evaluating multiple AI review systems through an experiment in which AI-generated reviews are shown only to authors after the author-response period \cite{emnlp2026experiment}.

\section{LLM Tools Used by Reviewers and ACs}
\label{sec:postsurveyadd}

Among the 662 reviewers who reported using an LLM, 636 (96.1\%) named at least one specific tool in the survey's free-text responses. ChatGPT was the most frequently named tool (53.1\% of tool mentions), followed by Claude (22.6\%) and Gemini (15.3\%). Among 333 tool mentions for which an access tier was specified, most referred to standard versions (59.2\%), with smaller shares referring to enterprise or university-licensed (16.2\%), pro/max/plus (16.5\%), and free (7.2\%) versions. \looseness=-1

Among the 75 ACs who reported using an LLM, 72 (96.0\%) named at least one specific tool in the survey's five free-text fields, yielding 86 tool entries. ChatGPT was the most frequently named tool (49 of 86 entries, 57.0\%), followed by Claude (18, 20.9\%) and Gemini (12, 14.0\%). The remaining 7 entries (8.1\%) referred to other tools, including DeepSeek, Grammarly, and Kimi, among others. Among 41 tool entries for which an access tier was specified, 39.0\% referred to standard versions, 39.0\% to enterprise or university-licensed versions, 14.6\% to pro/max/plus versions, and 7.3\% to free versions. \looseness=-1

For both reviewers and ACs, tool names and access tiers alone provided insufficient information to assess whether specific uses complied with the assigned policy and its confidentiality and privacy requirements. This limitation highlights the need for better mechanisms through which reviewers and ACs can disclose their LLM use. \looseness=-1

\section{Full Survey Instrument}
\label{app:survey}

In this section, we provide the full survey. Bracketed gray labels indicate
display logic: a question was shown only to respondents matching the label.
Circles ($\bigcirc$) denote single-select items and squares ($\square$) denote
multi-select items.

\subsection{Role \route{Everyone}}

\SQ{What role did you play in this year's ICML? Select all that apply. Note that
we make no distinction between main track and position track.}
\begin{checks}
  \item Author
  \item Reviewer (select this option if you provided at least one official paper
        review; ethics review or meta-review doesn't count)
  \item AC
  \item SAC
  \item None of the above
\end{checks}

\subsection{Understanding Reviewers' LLM Use \route{Reviewer}}

\SQ{Did you use LLM(s) at any point during the ICML 2026 review process?}
\begin{choices}
  \item Yes
  \item No
\end{choices}

\SQ{Why did you not use LLM(s) in your review? Please provide a brief description
(1--3 sentences). \route{No LLM Use}}
\freetext

\SQ{Which LLM(s) did you use in your reviews? Please list all of them. For each
LLM, please include its name (e.g., ChatGPT, Gemini, Claude), type (e.g., standard
subscription or enterprise/\allowbreak workplace/\allowbreak university-licensed), and version (e.g.,
GPT-5.5 with medium intelligence), to the extent you remember. \route{Yes LLM Use}}
\begin{itemize}[leftmargin=2em,itemsep=0.15em,topsep=0.25em]
  \item Name: \blank[4em]\quad Type: \blank[4em]\quad Version: \blank[4em]
  \item Name: \blank[4em]\quad Type: \blank[4em]\quad Version: \blank[4em]
  \item Name: \blank[4em]\quad Type: \blank[4em]\quad Version: \blank[4em]
\end{itemize}

\SQ{How did you use LLM(s) in your review? Please provide a brief description
(1--3 sentences). \route{Yes LLM Use}}
\freetext

\SQ{Please select all the ways you used LLM(s) in your review. \route{Yes LLM Use}}
\begin{checks}
  \item Understanding the paper (e.g., clarifying concepts or background)
  \item Checking technical correctness (e.g., sanity checks on experiments or math)
  \item Reviewing related work (e.g., verifying citations or identifying missing
        references)
  \item Identifying strengths or weaknesses of the paper
  \item Brainstorming feedback, questions, or edge cases
  \item Drafting an outline or parts of the review
  \item Expanding notes or bullet points into full text
  \item Editing or polishing reviewer-written text (for clarity, grammar, or tone)
  \item Synthesizing perspectives across reviews during the discussion phase
  \item Understanding, evaluating, or responding to the authors' rebuttal
  \item Other: \blank\blank
\end{checks}

\SQ{How carefully did you read and verify the LLM outputs? \route{Yes LLM Use}}
\begin{choices}
  \item Not carefully
  \item Slightly carefully
  \item Moderately carefully
  \item Very carefully
  \item Extremely carefully
\end{choices}

\SQ{How much do you stand behind the claims in your review(s) submitted to ICML 2026? \route{Yes LLM Use}}
\begin{choices}
  \item I would be hesitant to stand behind most claims in my review(s)
  \item I would be hesitant to stand behind several claims in my review(s)
  \item I would stand behind most claims, but be hesitant about a small number of
        claims in my review(s)
  \item I would stand behind all substantive claims in my review(s)
  \item I would stand behind all claims in my review(s), including minor details
\end{choices}

\SQ{How did using LLM(s) affect the following aspects of your review experience,
compared to not using LLM(s)? \route{Yes LLM Use; one option per row}}

\begin{center}
\begin{tabularx}{\columnwidth}{
    @{}
    >{\raggedright\arraybackslash}p{0.64\columnwidth}
    *{5}{>{\centering\arraybackslash}X}
    @{}
}
\toprule
 & $\downarrow\downarrow$ & $\downarrow$ & $-$ & $\uparrow$ & $\uparrow\uparrow$ \\
\midrule
Time spent                                  & $\bigcirc$ & $\bigcirc$ & $\bigcirc$ & $\bigcirc$ & $\bigcirc$ \\
Effort required                             & $\bigcirc$ & $\bigcirc$ & $\bigcirc$ & $\bigcirc$ & $\bigcirc$ \\
Review quality                              & $\bigcirc$ & $\bigcirc$ & $\bigcirc$ & $\bigcirc$ & $\bigcirc$ \\
Confidence in your evaluation               & $\bigcirc$ & $\bigcirc$ & $\bigcirc$ & $\bigcirc$ & $\bigcirc$ \\
Level of engagement in the review process  & $\bigcirc$ & $\bigcirc$ & $\bigcirc$ & $\bigcirc$ & $\bigcirc$ \\
Sense of ownership of your review           & $\bigcirc$ & $\bigcirc$ & $\bigcirc$ & $\bigcirc$ & $\bigcirc$ \\
\bottomrule
\end{tabularx}
\vspace{1pt}

$\downarrow\downarrow$: significantly decreased; 
$\downarrow$: slightly decreased; 
$-$: no change; \\
$\uparrow$: slightly increased; 
$\uparrow\uparrow$: significantly increased.
\end{center}

\subsection{Understanding ACs' LLM Use \route{AC}}

\SQ{Did you use LLM(s) at any point during the ICML 2026 review process to help
with your AC duties (e.g., to understand or summarize papers, reviews, or
discussion, or to draft meta-reviews)?}
\begin{choices}
  \item Yes
  \item No
\end{choices}

\SQ{Why did you not use LLM(s) in your AC duties? Please provide a brief
description (1--3 sentences). \route{No LLM Use}}
\freetext

\SQ{Which LLM(s) did you use to help with your AC duties? Please list all of them.
For each LLM, please include its name (e.g., ChatGPT, Gemini, Claude), type (e.g.,
standard subscription or enterprise/\allowbreak workplace/\allowbreak university-licensed), and version
(e.g., GPT-5.5 with medium intelligence), to the extent you remember.
\route{Yes LLM Use}}
\begin{itemize}[leftmargin=2em,itemsep=0.15em,topsep=0.25em]
  \item Name: \blank[4em]\quad Type: \blank[4em]\quad Version: \blank[4em]
  \item Name: \blank[4em]\quad Type: \blank[4em]\quad Version: \blank[4em]
  \item Name: \blank[4em]\quad Type: \blank[4em]\quad Version: \blank[4em]
\end{itemize}

\SQ{How did you use LLM(s) to help with your AC duties? Please provide a brief
description (1--3 sentences). \route{Yes LLM Use}}
\freetext

\SQ{How did using LLM(s) affect the following aspects of your AC experience,
compared to not using LLM(s)? \route{Yes LLM Use; one option per row}}

\begin{center}
\begin{tabularx}{\columnwidth}{
    @{}
    >{\raggedright\arraybackslash}p{0.64\columnwidth}
    *{5}{>{\centering\arraybackslash}X}
    @{}
}
\toprule
 & $\downarrow\downarrow$ & $\downarrow$ & $-$ & $\uparrow$ & $\uparrow\uparrow$ \\
\midrule
Time spent                                  & $\bigcirc$ & $\bigcirc$ & $\bigcirc$ & $\bigcirc$ & $\bigcirc$ \\
Effort required                             & $\bigcirc$ & $\bigcirc$ & $\bigcirc$ & $\bigcirc$ & $\bigcirc$ \\
Depth of interaction with reviewers                             & $\bigcirc$ & $\bigcirc$ & $\bigcirc$ & $\bigcirc$ & $\bigcirc$ \\
Meta-review quality                              & $\bigcirc$ & $\bigcirc$ & $\bigcirc$ & $\bigcirc$ & $\bigcirc$ \\
Confidence in your evaluation               & $\bigcirc$ & $\bigcirc$ & $\bigcirc$ & $\bigcirc$ & $\bigcirc$ \\
Level of engagement in AC duties  & $\bigcirc$ & $\bigcirc$ & $\bigcirc$ & $\bigcirc$ & $\bigcirc$ \\
Sense of ownership of your meta-review           & $\bigcirc$ & $\bigcirc$ & $\bigcirc$ & $\bigcirc$ & $\bigcirc$ \\
\bottomrule
\end{tabularx}
\vspace{1pt}

$\downarrow\downarrow$: significantly decreased; 
$\downarrow$: slightly decreased; 
$-$: no change; \\
$\uparrow$: slightly increased; 
$\uparrow\uparrow$: significantly increased.
\end{center}

\subsection{Future Policies on Reviewers' LLM Use \route{Everyone}}

\medskip\noindent
In this section, we are interested in your views on how LLMs should or should not
be used by reviewers. (If you have thoughts about the use of LLMs by ACs or SACs,
or other aspects of LLM use in peer review, please share them in the free-form
feedback section at the end of the survey.)

\SQ{Assuming all reviewers will fully follow any policy, which activities, if any,
do you think reviewers should be allowed to use LLMs for? Select all that apply.}
\begin{checks}
  \item Understanding the paper (e.g., clarifying concepts or background)
  \item Checking technical correctness (e.g., sanity checks on experiments or math)
  \item Reviewing related work (e.g., verifying citations or identifying missing
        references)
  \item Identifying strengths or weaknesses of the paper
  \item Brainstorming feedback, questions, or edge cases
  \item Drafting an outline or parts of the review
  \item Expanding notes or bullet points into full text
  \item Editing or polishing reviewer-written text (for clarity, grammar, or tone)
  \item Synthesizing perspectives across reviews during the discussion phase
  \item Understanding, evaluating, or responding to the authors' rebuttal
  \item Other: \blank\blank
  \item None of the above. All review activities should be done without LLM
        assistance
\end{checks}

\SQ{Assuming all reviewers will fully follow any policy, which aspects of a paper,
if any, do you think reviewers should be allowed to use LLMs to assess? Select all
that apply.}
\begin{checks}
  \item Soundness
  \item Presentation
  \item Significance
  \item Originality
  \item Limitations
  \item Ethical concerns
  \item Other: \blank\blank
  \item None of the above. All aspects should be assessed without LLM assistance
\end{checks}

\SQ{Now relaxing the assumption that all reviewers will fully follow any policy,
what would be your preferred policy around LLM use in peer review? Is it the same
as or different from your answer above? If different, please explain why (e.g.,
concerns about what is realistically possible to verify or enforce in practice).}
\freetext

\SQ{Should ICML consider piloting fully automated, clearly labeled LLM-generated
reviews as an additional input during the initial review stage, alongside human
reviews? (Note: Authors and Area Chairs would be explicitly informed that the
review was LLM-generated, similar to AAAI's 2026 pilot.)}
\begin{choices}
  \item Yes, for all submissions
  \item Yes, but only as an optional, opt-in service for authors
  \item No, ICML should not use fully automated reviewers
  \item Unsure, I would need more information before forming an opinion
\end{choices}

\subsection{Open-Ended Questions \route{Everyone}}

\SQ{Do you feel any of your own LLM use during this or other review cycles may
have been inappropriate? This survey is anonymous, and we value your honest
perspective. (Skip if you have not used LLM(s) for peer review.)}
\freetext

\SQ{Have you observed or heard about any specific instances of potentially
inappropriate LLM use in peer review? (Please exclude cases based solely on
general suspicion.) How do you think such cases should be detected and addressed?}
\freetext

\SQ{What practices can help reviewers produce high-quality reviews and maintain a
strong sense of ownership over their reviews while using LLMs?}
\freetext

\subsection{Participant Background \route{Everyone}}

\SQ{Which review policy were you assigned for ICML 2026? \route{Reviewer}}
\begin{choices}
  \item Policy A (Conservative)
  \item Policy B (Permissive)
  \item I'm not sure
\end{choices}

\SQ{Did you receive a reviewer recognition award at ICML 2026? \route{Reviewer}}
\begin{choices}
  \item Gold reviewer
  \item Silver reviewer
  \item No
  \item I'm not sure
\end{choices}

\SQ{What is your current position?}
\begin{choices}
  \item Undergraduate student
  \item Master's student
  \item PhD student
  \item Postdoctoral researcher
  \item Faculty member
  \item Industry researcher
  \item Other: \blank\blank
\end{choices}

\SQ{If you have received a PhD, when did you receive it?}
\begin{choices}
  \item I have not received a PhD
  \item 2024--2026
  \item 2021--2023
  \item 2020 or earlier
\end{choices}

\SQ{Approximately how many times have you served as a paper reviewer for ICML,
NeurIPS, or ICLR, including this ICML 2026 review cycle? (Exclude service as an
ethics reviewer, AC, or SAC.) \route{Reviewer}}
\begin{choices}
  \item 1 review cycle
  \item 2--3 review cycles
  \item 4--6 review cycles
  \item 7--10 review cycles
  \item More than 10 review cycles
\end{choices}

\SQ{Approximately how many times have you served as an AC for ICML, NeurIPS, or
ICLR, including this ICML 2026 review cycle? (Exclude service as a SAC.)
\route{AC}}
\begin{choices}
  \item 1 review cycle
  \item 2--3 review cycles
  \item 4--6 review cycles
  \item 7--10 review cycles
  \item More than 10 review cycles
\end{choices}

\SQ{Have you spent three or more years at an institution where English was the
primary working language? \route{Everyone}}
\begin{choices}
  \item Yes
  \item No
\end{choices}

\subsection{End \route{Everyone}}

\SQ{Please use this space to share any additional thoughts on the use of LLMs in
peer review, this year's policy, or this survey. We value your feedback!}
\freetext

\section{ICML 2026 Main-Track Review Form}
\label{sec:reviewform}

In this section, we describe the review form that reviewers completed for main-track papers. All fields were required unless marked optional. See \url{https://icml.cc/Conferences/2026/ReviewerInstructions#MainTrack} for full details.

\subsection*{Summary}

Briefly summarize the paper and its contributions. This is not the place to critique the paper; the authors should generally agree with a well-written summary. This is also not the place to paste the abstract; please provide the summary in your own understanding after reading.

\noindent\framebox[\linewidth][l]{\rule{0pt}{3em}}

\subsection*{Strengths and Weaknesses}

Please provide a thorough assessment of the strengths and weaknesses of the paper, touching on each of the following dimensions: soundness, presentation, significance, and originality. \textbf{We encourage you to be open-minded about the potential strengths and broad definitions of significance and originality.} For example, originality may arise from creative combinations of existing ideas, application to a real-world use case, or removing restrictive assumptions from prior theoretical results. We provide detailed guidelines below on each dimension:

\begin{itemize}
    \item \textit{Soundness}: Is the submission technically sound? Are claims well supported (e.g., by theoretical analysis or experimental results)? Are the methods used appropriate? If the paper includes theoretical results, are the proofs correct and based on reasonable assumptions? If the paper includes empirical results, are the experiments well-designed? Are the authors careful and honest about evaluating both the strengths and weaknesses of their work? 
    \textit{Note: Soundness is distinct from impact. A paper can be technically sound—meaning correct, rigorous, and methodologically appropriate—even if its contributions are modest or incremental. Conversely, a paper proposing a high-impact idea must still meet the same bar for technical soundness. Reviewers should assess these dimensions separately.}
    \item \textit{Presentation}: Is the submission clearly written and well structured? (If not, please make constructive suggestions for improving its clarity.) Is the overall narrative easy to follow? Does the work properly position itself in the context of prior/concurrent literature and clearly discuss how it differs? (Note that a superbly written paper provides enough information for an expert reader to reproduce its results.)
    \item \textit{Significance}: Does the paper address an important or relevant problem? Does it advance understanding, capabilities, or practice in machine learning? Could it influence future research or applications (e.g., other researchers or practitioners are likely to use the ideas or build on them)? Is the scope of impact broad or specialized, and is that appropriate for the contribution? Even if the improvements are modest or domain-specific, could they unlock new directions or provide practical utility?
    \item \textit{Originality}: Does the work provide new insights, deepen understanding, or highlight important properties of existing methods? Does the work introduce new tasks, methods, theory, data, or perspectives that advance the field in some dimensions? Does this work offer a novel combination of existing techniques, and is the reasoning behind this combination well-articulated? Are the contributions clearly distinguished from closely related literature, and is the novelty well justified? \textbf{As the questions above indicates, originality does not necessarily require introducing an entirely new method. Rather, a work that provides novel insights by evaluating existing methods, or demonstrates improved understanding is also equally valuable.}
\end{itemize}

\noindent\framebox[\linewidth][l]{\rule{0pt}{3em}}

\subsection*{Ratings}

Based on what you discussed in “Strengths and Weaknesses”, please rate the paper on the following scale to indicate the significance of the overall contribution this paper makes to the research area being studied. If you select “fair” or “poor” (indicating that the paper falls short of the standard), ensure that “Strengths and Weaknesses” include a clear justification of your rating. 

\begin{center} \begin{tabular}{ @{\hspace{0em}} l @{\hspace{1.2em}} l @{\hspace{1.2em}} l @{\hspace{1.2em}} l @{\hspace{0em}} } \textbf{Soundness} & \textbf{Presentation} & \textbf{Significance} & \textbf{Originality} \\[2pt] $\bigcirc$ 4: excellent & $\bigcirc$ 4: excellent & $\bigcirc$ 4: excellent & $\bigcirc$ 4: excellent \\ $\bigcirc$ 3: good & $\bigcirc$ 3: good & $\bigcirc$ 3: good & $\bigcirc$ 3: good \\ $\bigcirc$ 2: fair & $\bigcirc$ 2: fair & $\bigcirc$ 2: fair & $\bigcirc$ 2: fair \\ $\bigcirc$ 1: poor & $\bigcirc$ 1: poor & $\bigcirc$ 1: poor & $\bigcirc$ 1: poor \end{tabular} \end{center}

\subsection*{Key Questions for Authors}

If you have any \textit{important} questions for the authors, please carefully formulate them here (ideally around 3-5). Please reserve your questions for cases where the response would likely change your evaluation of the paper, clarify a point in the paper that you found confusing, or address a critical limitation you identified. Please number your questions so authors can easily refer to them in the response, and explain how possible responses would change your evaluation of the paper.

\noindent\framebox[\linewidth][l]{\rule{0pt}{3em}}

\subsection*{Limitations}

Have the authors adequately discussed the limitations and potential negative societal impact of their work? If so, simply enter “yes”; if not, please include constructive suggestions for improvement. \textit{Authors should be rewarded rather than punished for being up front about the limitations of their work and any potential negative societal impact.}

\noindent\framebox[\linewidth][l]{\rule{0pt}{3em}}

\subsection*{Overall Recommendation}
Please provide an overall recommendation for this submission. Choices:

\begin{choices}
    \item \textbf{6: Strong Accept}: Technically flawless paper with exceptional impact on one or more areas of AI, with strong evaluation, reproducibility, and resources, and no unaddressed ethical considerations.
    \item \textbf{5: Accept}: Technically solid paper, with high impact on at least one sub-area of AI or moderate-to-high impact on more than one area of AI, with good-to-excellent evaluation, resources, reproducibility, and no unaddressed ethical considerations.
    \item \textbf{4: Weak Accept}: Technically solid paper that advances at least one sub-area of AI, with a contribution that others are likely to build on, but with some weaknesses that limit its impact (e.g., limited evaluation). Please use sparingly.
    \item \textbf{3: Weak Reject}: A paper with clear merits, but also some weaknesses, which overall outweigh the merits. Papers in this category require revisions before they can be meaningfully built upon by others. Please use sparingly.
    \item \textbf{2: Reject}: For instance, a paper with technical flaws, weak evaluation, inadequate reproducibility, incompletely addressed ethical considerations, or writing so poor that it is not possible to understand its key claims.
    \item \textbf{1: Strong Reject}: For instance, a paper with well-known results, unaddressed ethical considerations, or a poorly written paper where it is impossible to tell what the nature of its contribution is.
\end{choices}

\subsection*{Confidence}
Please provide a "confidence score" for your assessment of this submission to indicate how confident you are in your evaluation. Choices:

\begin{choices}
    \item \textbf{5}: You are absolutely certain about your assessment. You are very familiar with the related work and checked the math/other details carefully.
    \item \textbf{4}: You are confident in your assessment, but not absolutely certain. It is unlikely, but not impossible, that you did not understand some parts of the submission or that you are unfamiliar with some pieces of related work.
    \item \textbf{3}: You are fairly confident in your assessment. It is possible that you did not understand some parts of the submission or that you are unfamiliar with some pieces of related work. Math/other details were not carefully checked.
    \item \textbf{2}: You are willing to defend your assessment, but it is quite likely that you did not understand the central parts of the submission or that you are unfamiliar with some pieces of related work. Math/other details were not carefully checked.
    \item \textbf{1}: Your assessment is an educated guess. The submission is not in your area, or the submission was difficult to understand. Math/other details were not carefully checked.
\end{choices}

\subsection*{Ethical Review Flag (Optional)}

If you believe there are ethical issues with this paper, please flag the paper for an ethics review. For guidance on when this is appropriate, please review the ICML research ethics guidelines.
\begin{checks}
    \item Flag this paper for an ethics review.
\end{checks}

\subsection*{Ethics Expertise Needed (Optional)}

If you flagged this paper for ethics review, what area of expertise would it be most useful for the ethics reviewer to have? Please click all that apply: 

\begin{checks}
    \item Discrimination / Bias / Fairness Concerns
    \item Inappropriate Potential Applications \& Impact (e.g., human rights concerns)
    \item Responsible Research Practice (e.g., IRB, documentation, research ethics)
    \item Privacy and Security (e.g., personally identifiable information)
    \item Legal Compliance (e.g., EU AI Act, GDPR, copyright, terms of use)
    \item Research Integrity Issues (e.g., plagiarism)
    \item Other
\end{checks}

\subsection*{Ethical Review Concerns (Optional)}

If you flagged this paper for an ethics review, please explain your concerns in detail.

\noindent\framebox[\linewidth][l]{\rule{0pt}{3em}}

\subsection*{Compliance with LLM Reviewing Policy}

Please affirm the following: \textit{While writing reviews, I have fully complied with the ICML 2026 policy for LLM use in reviewing. I confirm that I have strictly followed the \textbf{actual policy} assigned to me, as displayed on my Reviewer Console. I understand that any deviation from this assigned policy constitutes a violation of the conference's academic integrity guidelines and can lead to desk rejection of my own submissions.}
\begin{checks}
    \item Affirmed.
\end{checks}

\subsection*{Code of Conduct Acknowledgement}

Please affirm the following: \textit{While performing my duties as a reviewer (including writing reviews and participating in discussions), I have and will continue to abide by the ICML code of conduct.}
\begin{checks}
    \item Affirmed.
\end{checks}

\subsection*{Final Justification (Post-Rebuttal)}

Please explain your final recommendation, taking into account both the paper and the authors’ rebuttal. Summarize how you weighed the strengths and weaknesses across dimensions such as soundness, originality, significance, and clarity. Importantly, \textbf{please indicate whether the rebuttal addressed your main concerns, changed your evaluation, or reinforced your prior assessment.} This justification is shared with the authors, AC, SAC, and PCs, so ensure it is constructive, respectful, and clearly communicates your final position.

\noindent\framebox[\linewidth][l]{\rule{0pt}{3em}}

\section{ICML 2026 AC Review Quality Rating}
\label{sec:acreviewquality}

Below is the question shown to ACs to assess the quality of each review.

\subsection*{Reviewer Rating}

Please refer to the reviewer guidelines: \url{https://icml.cc/Conferences/2026/ReviewerInstructions}

\begin{choices}
    \item 0: EXTREMELY LOW -- a generic review applicable to any paper / a short and dismissive review / a review that appears to be fully LLM-generated.
    \item 1: VERY LOW -- a review with serious flaws on multiple aspects.
    \item 2: SOMEWHAT LOW -- a review with serious flaws on one aspect / a review without serious flaws but with limited insights for authors/ACs, reviewer not sufficiently responsive to authors/ACs.
    \item 3: AVERAGE -- acceptable review, but nothing stands out; moderately helpful for decision-making; reviewer responsive to authors/ACs.
    \item 4: SOMEWHAT GOOD -- a helpful review that stands out on some aspects and provides useful insights; reviewer responsive to authors/ACs.
    \item 5: VERY GOOD -- a very insightful review that stands out on all aspects; reviewer responsive to authors/ACs.
    \item 6: EXTREMELY GOOD -- an excellent review that helps authors to non-trivially improve the paper / brings a unique piece of information that is crucial for the decision; reviewer responsive to authors/ACs.
\end{choices}

\subsection*{Rating Justification (Optional)}

Optional justification of your rating.

\noindent\framebox[\linewidth][l]{\rule{0pt}{3em}}


\begin{figure*}[b]
    \centering
    \includegraphics[width=\linewidth]{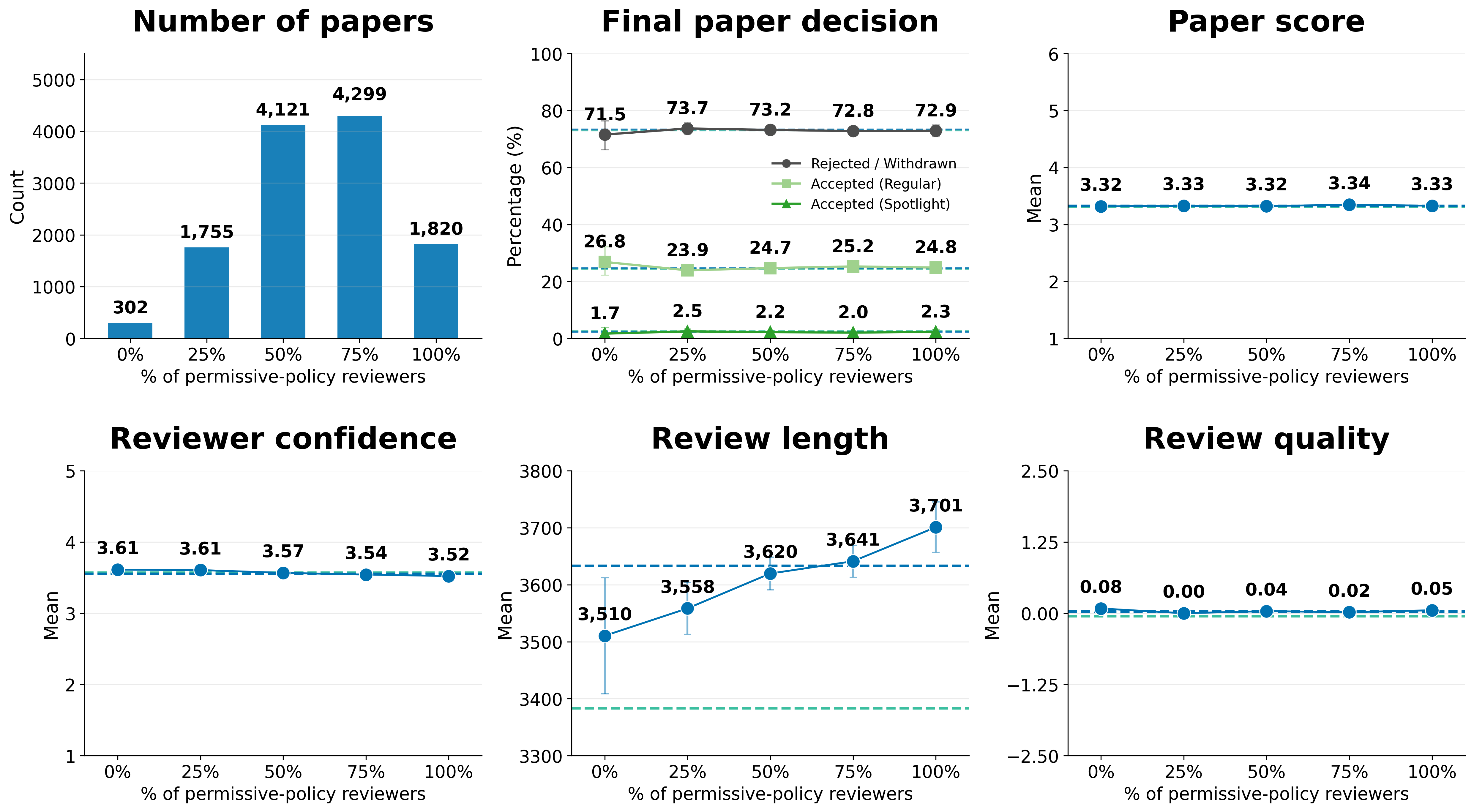}
    \caption{\textbf{Outcomes among permissive-policy papers by share of reviewers assigned to the permissive policy.} Analyses include the 12,297 permissive-policy papers (80.9\%) with exactly four reviewers whose policy assignments were known, yielding compositions of 0\%, 25\%, 50\%, 75\%, or 100\% permissive-policy reviewers. Papers with other reviewer counts or insufficient assignment information are excluded. The first panel shows the number of papers in each composition; the remaining panels show corresponding review outcomes. Points indicate decision percentages or paper-level means, with 95\% confidence intervals. Dashed lines show the corresponding overall values for all P$\rightarrow$C (teal) and P (blue) papers in \cref{fig:experiment1}. The similarity of outcomes across reviewer compositions suggests that the main results are not driven by differences in the mix of reviewers assigned to the permissive policy.}
    \label{fig:experiment1_robustness}
\end{figure*}

\begin{figure*}[b] 
\centering 
\includegraphics[width=\linewidth]{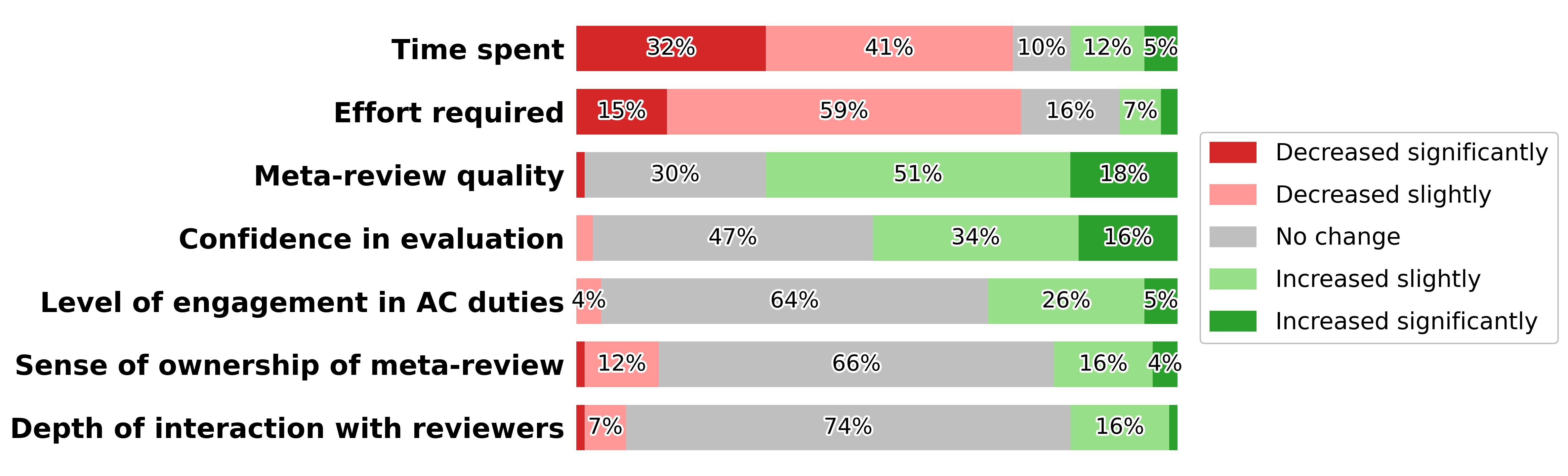} 
\caption{\textbf{Post-survey results on ACs' self-reported effects of LLM use} (N=73). ACs reported patterns broadly similar to those observed among reviewers (\cref{fig:reviewer-perceptions}), including improved meta-review quality, reduced time spent and effort, and increased confidence. Effects on engagement in AC duties, ownership over meta-reviews, and depth of interaction with reviewers were more mixed, with no change the most common response.}
\label{fig:ac-perceptions}
\end{figure*}

\begin{figure*}[t]
    \centering
    \includegraphics[width=\textwidth]{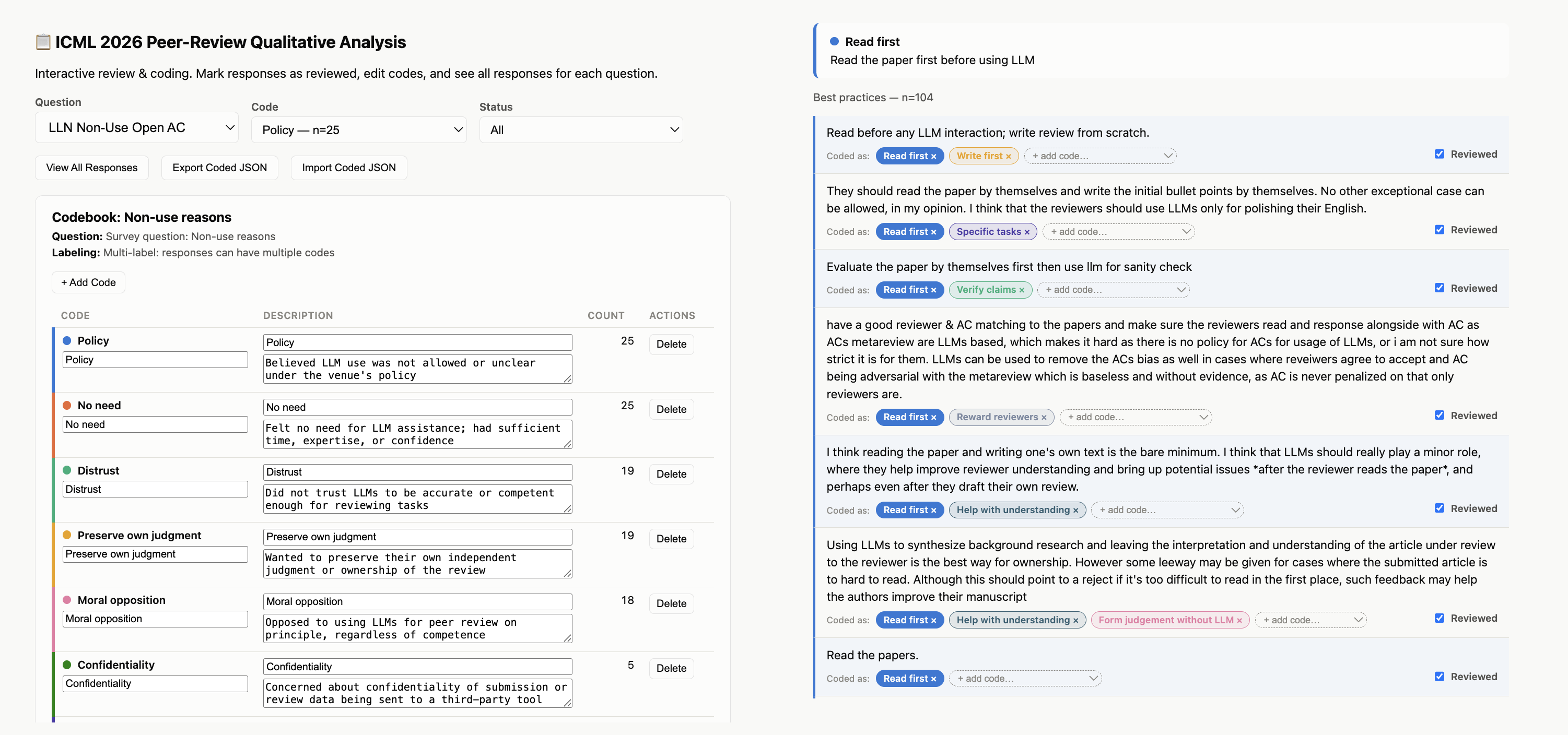}
    \caption{
        \textbf{Qualitative coding interface for the post-survey.}
        \textbf{(Left)} The codebook view supports selecting a survey question
        and filtering responses by code and review status. The illustrated
        ``Non-use reasons'' codebook displays color-coded entries, editable
        code names and descriptions, and response counts, alongside controls
        for adding or deleting codes and importing or exporting coded data.
        \textbf{(Right)} The response view displays quotations assigned the
        ``Read first'' code within ``Best practices,'' together with its
        definition (104 matching responses; a subset is visible). Each
        quotation shows the selected code and any co-occurring codes, such
        as ``Write first'' or ``Verify claims.'' Codes can be added or removed
        for individual responses, supporting multi-label coding. 
    }
    \Description{
        Two side-by-side screenshots of a qualitative coding interface.
        The left panel shows question, code, and status filters above an
        editable codebook. The right panel shows survey responses tagged
        Read first, additional code tags, and checked Reviewed boxes.
    }
    \label{fig:qualitative-coding-interface}
\end{figure*}

\begin{figure*}[b] 
\centering 
\fbox{\includegraphics[width=0.7\linewidth]{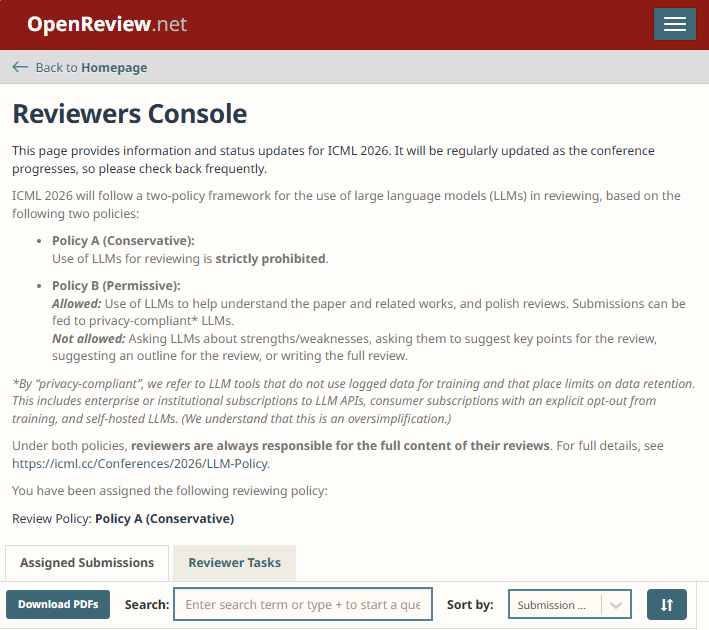}}
\caption{\textbf{Screenshot of the OpenReview reviewer console showing the assigned policy.} At ICML 2026, the conservative and permissive policies were called ``Policy A'' and ``Policy B,'' respectively. We omit these labels in the paper to reduce confusion. \looseness=-1}
\label{fig:reviewerconsole}
\end{figure*}

\begin{figure*}[b] 
\fbox{\includegraphics[width=\linewidth]{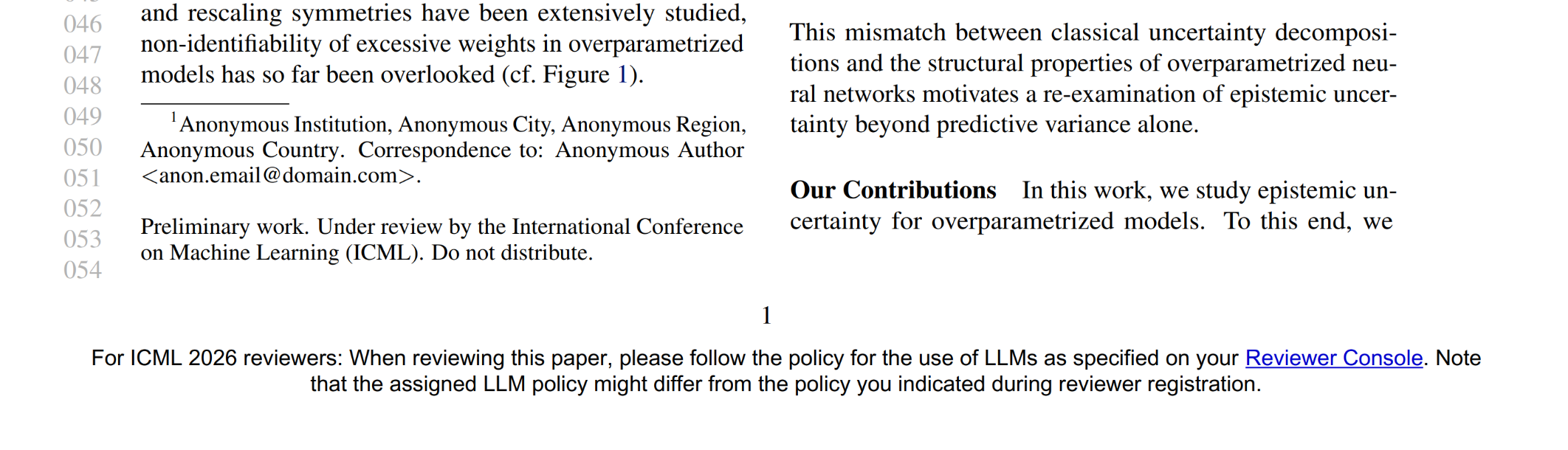}}
\caption{\textbf{Screenshot showing the policy reminder displayed at the bottom of the first page of every submission.}}
\label{fig:reminder}
\end{figure*}

\end{document}